\documentclass[twocolumn]{aastex62}
\usepackage{hyperref}
\usepackage{placeins}
\pdfoutput=1 %for arXiv submission
\usepackage{amsmath,amstext}
\usepackage[T1]{fontenc}
\usepackage{ae,aecompl}
\usepackage[utf8x]{inputenc}
\usepackage{atbegshi}
\usepackage{apjfonts} 
\usepackage{graphics}
\usepackage[figure,figure*]{hypcap}
\usepackage{enumitem}
\usepackage{url}
\usepackage{bm}
\usepackage{lineno}
\usepackage[normalem]{ulem}
\input{hyperlink-year-only-natbib-patch}
\usepackage{float}

\usepackage{lipsum}
\graphicspath{{./}{figures/}}
\newcommand{\rsp}{$R_{sp}$}

\newcommand{\ts}[1]{\textcolor{olive}{#1}}

\def\redh{\rm{Red_{high}}}
\def\redl{\rm{Red_{low}}}
\usepackage[T1]{fontenc} % if needed

\begin{document}

\title{Connecting galaxy evolution in clusters with their radial profiles and phase space distribution: results from the IllustrisTNG hydrodynamical simulations}
%The signatures of galaxy evolution on the radial distribution of galaxies and the splashback radius within cluster mass halos in IllustrisTNG}

\author[0000-0002-4746-2128]{Tara Dacunha}
\affiliation{Department of Physics and Astronomy, University of Pennsylvania, 209 S. 33rd St. Philadelphia, PA 19104, USA}

\author[0000-0003-4778-6170]{Matthew Belyakov}
\affiliation{Department of Physics and Astronomy, University of Pennsylvania, 209 S. 33rd St. Philadelphia, PA 19104, USA}
\affiliation{Division of Geological and Planetary Sciences, California Institute of Technology, Pasadena, CA 91125, USA}
\author[0000-0002-0298-4432]{Susmita Adhikari}
\affiliation{Department of Astronomy and Astrophysics, University of Chicago, Chicago, IL 60637, USA}
\affiliation{Kavli Institute for Cosmological Physics, University of Chicago, Chicago, IL 60637, USA}
\author[0000-0002-6389-5409]{Tae-hyeon Shin}
\affiliation{Department of Physics and Astronomy, Stony Brook University,Stony Brook, NY 11794, USA}
\author[0000-0003-3155-245X]{Samuel Goldstein}
\affiliation{Department of Physics and Astronomy, University of Pennsylvania, 209 S. 33rd St. Philadelphia, PA 19104, USA}
\affiliation{Max Planck Institute for Astrophysics, Karl-Schwarzschild-Strasse 1, 85748 Garching, Germany}
\author[0000-0002-8220-3973]{Bhuvnesh Jain}
\affiliation{Department of Physics and Astronomy, University of Pennsylvania, 209 S. 33rd St. Philadelphia, PA 19104, USA}
%\affiliation{University of Pennsylvania, Department of Astronomy}

\begin{abstract}

We study the population of galaxies around galaxy clusters in the hydrodynamic simulation suite IllustrisTNG 300-1 to study the signatures of their evolutionary history on observable properties. We measure the  radial number density profile, phase space distribution, and  splashback radius for galaxies of different  masses and colors over the redshift range $z=0-1$. The three primary physical effects which shape the galaxy distribution within clusters are the galaxy quenching, angular momentum distribution and  dynamical friction. We find three distinct populations of galaxies by applying a Gaussian mixture model to their distribution in color and mass. They have distinct evolutionary histories and leave distinct signatures on their distribution around cluster halos.  We find that low-mass red galaxies show the most concentrated distribution in clusters and the largest splashback radius, while  high-mass red galaxies show a less concentrated distribution and a smaller splashback radius. Blue galaxies, which mostly quench into the low-mass red population, have the shallowest distribution within the clusters, with those on radial orbits quenched rapidly before reaching pericenter. Comparison with the distribution of galaxies from the Dark Energy Survey (DES) survey around Sunyaev-Zeldovich (SZ) clusters from the Atacama Cosmology Telescope (ACT) and South Pole Telescope (SPT) surveys shows evidence for differences in galaxy evolution between simulations and data.

%we find that the three leave signatures on thlow total mass and high total mass red galaxies which is reflected in their evolutionary histories, star formation rates, and splashback signatures. Comparing to data from the Dark Energy Survey (DES), we find that Blue galaxies in IllustrisTNG have a splashback signal absent in data, and that the low-mass Red population in IllustrisTNG more closely matches Red galaxies in data than the higher mass Red population. \td{change preceding sentence to address our color-mass populations in splashback, then data comparison, and then dynamical friction}
\end{abstract}

\section{Introduction:}
\label{Sec:Intro}

Galaxies evolve within the potential wells of massive dark matter halos, some of the largest of these bound structures are known as galaxy clusters. These are relatively easy to observe and characterize and provide an excellent laboratory to test the predictions of hierarchical structure formation and galaxy evolution. 

It has been known that galaxies that live in these dense environments are often elliptical galaxies that appear to have little or no active star-formation \citep{Oemler:1974yw, Dressler:1980wq,  Dressler:1984kh,  Balogh:1997bw, Poggianti:1999xh} . Various physical processes have been invoked to explain this galaxy star-formation/morphology-density relationship. This phenomenon is primarily expected to arise from astrophysical processes within the galaxy cluster that remove the reservoirs of cold gas from infalling galaxies and prevent them from forming stars. Various processes, with graphic names, like strangulation \citep{Larson:1980mv}, ram-pressure stripping \citep{GunnGott72}, and harassment \citep{Abadi:1999qy} of galaxies falling into the hot intra-cluster environment cause the star-formation rates of galaxies to decrease over time as they orbit through a halo’s potential well.

Recently, measurements of the splashback radius (\rsp) have provided new insight into understanding the evolution of collapsed structures like dark matter halos in the universe \citep{More:2016vgs, Baxter:2017csy,Chang:2017hjt,Shin19, Zuercher:2018prq,Murata:2020enz, adhikari20, Shin_2021}. The splashback feature appears at the edge of dark matter halos, separating the inner virialized, or multistreaming regions, from the infall regions, forming the boundary of dark matter halos \citep{Diemer:2014xya, Adhikari:2014lna, More:2015ufa, Shi:2016lwp}. The splashback surface forms near the first apocenter of the most recently infalling material. In idealized models of halo collapse there is a density caustic at splashback \citep{FG84, LD10, Adhikari:2014lna}, but in stacked clusters it manifests itself as a steepening of the slope of the density profile in a narrow localized region, where the inner halo term sharply falls off, giving away to a shallower infall density. Since the radial velocity of dark matter particles or satellite galaxies is low near the turnaround radius, they cluster at $R_{sp}$ \citep{Adhikari:2014lna}. For an observed galaxy cluster, $R_{sp}$ is often defined as the radius where the logarithmic slope of the galaxy number density profile or the dark matter density reaches a minimum (maximum steepness). %The splashback feature can therefore be defined by the location of the steepening in the density profile at the edge of a cluster. Specifically, the minimum of the logarithmic slope of the cluster profile gives the location of the splashback feature. %This is a result of the behavior of the density profile changing between the virialized region of the cluster and the infall region. 
%The virialized region follows a Navarro Frenk and White (NFW)-like profile with a slope of $-3$, and the infall section of the profile follows a power law with index around $-1.5$ \cite{Navarro_1996, Baxter_2017}. As a result, there is a sharp transition between the virialized and infalling regimes at the location of the splashback feature, such that the slope reaches a minimum at the location 2of the change from NFW to power-law, after which it returns near the background value.

As the splashback radius traces the first apocenter, it can also directly be associated with a timescale, i.e. the time taken for a galaxy or a dark matter particle to make one passage through a pericenter and reach its orbital apocenter from first infall \citep{adhikari20}. Therefore, if a population of galaxies shows a splashback-like feature, we can associate an average timescale to it. As intra-cluster processes that influence the evolution of galaxies are naturally sensitive to the time spent by a galaxy within the cluster potential it is important to understand how galaxy properties and their distribution in observed clusters can inform us about different cluster processes.

In this paper we investigate the connection between galaxy evolution and the colors of observed galaxies and study how different evolutionary histories leave their imprint on the phase space distributions and the shape of the full radial profile of a population of galaxies. We use the state-of-the-art IllustrisTNG \citep{Nelson2018} suite of hydrodynamic simulations for this analysis. \cite{Donnari_2020} analyzed quenching mechanisms in IllustrisTNG 100-1 and 300-1. They find that galaxies in IllustrisTNG show primarily two quenching pathways, determining that low-mass galaxies experience strong amounts of environmental quenching, while high-mass galaxies go through mass quenching driven by AGN and mergers. We study how these pathways leave signatures in the cluster field. \cite{neil21} have characterized the splashback radius of galaxies and dark matter in clusters in IllustrisTNG as a function of halo properties. In this work we study in detail how the distribution of galaxies depends on the galaxy star-formation properties and explore how we can use these distributions to constrain models of galaxy quenching in observations.  We complement and expand upon the results from \cite{Shin19} and \cite{adhikari20} who have examined the galaxy color and splashback connection by forward modelling observations from cold dark matter (CDM) N-body simulations.

In particular, the assembly history of galaxies in a cluster mass halo is imprinted on their spatial and velocity distribution within the cluster. The overall radial number density of galaxies is sensitive to various processes like tidal disruption, dynamical friction, and quenching, each of which are sensitive to the angular momentum and infall time of galaxies into the cluster. Moreover, the relative importance of the different processes can change as a function of halo properties such as accretion rate, redshift, and cluster age. Most observational studies of the splashback radius compare with results from CDM-N-body simulations, assigning galaxies to subhalos, in this paper we attempt to marginalize over the uncertainty in the subhalo-galaxy connection and compare directly to hydrodynamic simulations. While the details of the feedback model in hydrodynamic simulations may still be uncertain, it is ideal for studying the overall expected trends with galaxy properties. Comparison with data can then be helpful for both constraining the models in theory space and also for data interpretation. In this paper we primarily focus on galaxy populations separated by their color and magnitude properties, to understand how these different populations inhabit clusters.

This paper is organized as follows. In \autoref{Sec:Simulation}, we briefly describe the IllustrisTNG simulation and the halo and galaxy selection method we use. In \autoref{Sec:Properties}, we describe our color-mass population split and how we determine its location, discuss the differences in evolutionary history of color and mass, as well as the star-formation rate of the populations. In \autoref{Sec:PhaseSpace}, we compare the phase space distributions of the populations, emphasizing the distribution of gas, dark matter, and total mass in phase space. In \autoref{Sec:Profiles}, we first compare the galaxy number density profiles and resulting splashback feature for each of the three different color-mass populations over redshift. We then make comparisons to splashback results in data from \citep{Shin19} and \citep{adhikari20}. We also examine the potential for dynamical friction as an explanation for the differences in the splashback feature between populations. In \autoref{Sec:Conclusion} we summarize our results, compare to other work, propose observables, and discuss further directions.

\section{Simulation}
\label{Sec:Simulation}
%\subsection{Simulation Data}
We conduct our analysis on the IllustrisTNG 300-1 data set as presented in \cite{Nelson2018}; \cite{VolkerSpringel2018}; \cite{Marinacci2018}; \cite{Pillepich2018}; \cite{Naiman2018}. IllustrisTNG is a suite of large volume cosmological magnetohydrodynamic simulations which are based on an improved version of the original Illustris galaxy formation model and rely on the moving-mesh code \texttt{AREPO} \citep{Springel_2010, Nelson2018, Weinberger_2020}. All IllustrisTNG simulation runs use cosmological parameters from the Planck Collaboration: $\Omega_m = \Omega_{dm}+ \Omega_b= 0.3089, \Omega_b=0.0486, \Omega_\Lambda=0.6911, \sigma_8 = 0.8159, n_s=  0.9667$, with a Hubble  constant of $H_0=100 h \text{ km } \text{ s}^{-1} \text{ Mpc}^{-1},$ where $h=0.6774$ \citep{Planck18}. 

%The IllustrisTNG project consists of 18 separate runs, split into full hydrodynamic  and dark matter (DM) only simulations. In each category there are three different box sizes, with $50, 100, 300 ~h^{-1}\text{Mpc}$ boxes, each of which is then split into low, medium, and high resolution runs. Each run contains 99 snapshots, going from $z=20$ to $z=0$.

We use the TNG300-1 high-resolution simulation. The TNG300 simulation is run within a periodic cube with a side length of 302 Mpc, with TNG300-1 being the highest resolution box. TNG300-1 has an initial $2 \times 2500^3$ cells/particles, with a gas cell mass of $1.1\times 10^{7} h^{-1} M_{\odot}$, dark matter cell mass of $4.0\times 10^{7} h^{-1} M_{\odot}$, and baryonic matter cell mass of $7.6\times 10^{6} h^{-1} M_{\odot}$. Dark matter particles in the simulation have a gravitation softening length of 1.5 kpc in physical units for $z\leq1$ and comoving units for $z>1$ \citep{nelson2021illustristng}. The public release provides 99 snaphsots between $z=20$ to $z=0$. The large volume and high resolution make it ideal to study a large dynamic range of halo masses from cluster mass objects to their satellite galaxies.

%As compared to data from the Dark Energy Survey (DES) and the Atacama Cosmology Telescope (ACT) \citep{Shin19, adhikari20}, IllustrisTNG allows us to examine a smaller-mass galaxy and cluster sample. With future sky surveys such as Legacy Survey of Space and Time (LSST) at the Vera Rubin Observatory giving ability to see galaxies and clusters that were previously too dim to study, using simulations such as IllustrisTNG is ideal for planning avenues of exploration with new results. The IllustrisTNG 300-1 simulation currently has the largest number of galaxies resolved of currently existing cosmological simulations, at over $10^5$ galaxies with mass greater than $10^9 M_\odot$ \citep{nelson2021illustristng}.

\begin{table*}
    \begin{tabular}{|p{2cm}||p{1.2cm}|p{1.2cm}|p{1.2cm}|p{1.2cm}|p{1.2cm}|p{1.2cm}|p{1.2cm}|p{1.2cm}|p{1.2cm}|p{1.2cm}|}
        \hline
        Redshift & $0$  & $0.1$ & $0.2$ & $0.3$ & $0.4$ & $0.5$ & $0.6$ & $0.7$  & $0.8$ & $1.0$\\
        \hline
        \hline
        Halos & 684 & 599 & 530 & 454 & 390 & 311 & 279 & 214 & 169 & 118\\
        \hline 
        Blue Galaxies &116692 & 116429 & 113127 & 107374 & 99527 & 86337 & 80581 & 67246 & 56463 & 42705\\
        \hline
        Red$_{high}$ Galaxies & 36400 & 29962 & 24482 & 19866 & 15649 & 10932 & 9215 & 6648 & 4778 & 2841\\
        \hline
        Red$_{low}$ Galaxies & 199446 & 160292 & 128988 & 102991 & 80174 & 56055 & 46447 & 33062 & 23465 & 13026\\
        \hline
    \end{tabular}
    \caption{The number of halos with mass greater than $5\times 10^{13} h^{-1} M_{\odot}$ and galaxies of each color-mass population that are within $7\cdot R_{200m}$ of a massive halo. The galaxy numbers include double-counted galaxies, since a galaxy can be within $7\cdot R_{200m}$ of multiple massive halos. The proportion of both high mass and low-mass Red galaxies to total number of galaxies, increases at later times which is in line with the notion of quenching.}
    \label{tab:Table1} 
\end{table*}

\subsection{ Cluster and Galaxy Selection}
\label{Sec:selection}
We select ten snapshots from TNG300-1 between $z=0$ and $z=1$. Gravitationally bound structures are located by the simulation using the \texttt{SUBFIND} algorithm \citep{Springel_2001, Dolag_2009}. These structures are subsequently linked with each other and nearby particles using a Friends-of-Friends (FoF) algorithm \citep{davisfrenk}.
%In each FoF group, the most massive gravitationally bound object is labelled as the main halo, or the central galaxy, while other objects are labelled as subhalos, or satellite galaxies. The center of the halo is defined as the position of the most bound gravitationally particle within the main halo, as determined by the SUBFIND algorithm used in IllustrisTNG \citep{nelson2021illustristng}.

Our cluster  sample consists of halos with $M_{200m} > 5\times 10^{13} h^{-1} M_{\odot}$. Note that this is  smaller than the mass threshold of clusters typically used in the splashback analysis of observational data; however, we use this threshold to give us good statistics in the IllustrisTNG-300 box which has a significantly smaller volume than galaxy surveys like the Dark Energy Survey(DES) \citep{Abbott:2005bi} and The Hyper Suprime Cam (HSC) survey  \citep{2018PASJ...70S...4A}, and therefore fewer massive clusters above the mass threshold used in data.  We examine these populations in IllustrisTNG over different redshifts, but we primarily focus on the TNG300-1 snapshot 66 at redshift, $z=0.52$.  Our cluster sample consists of 311 halos with a mean halo mass of  $M_{200m} = 10^{13.9} h^{-1} M_{\odot}$. We make this choice of redshift to compare with observational results from \cite{Shin19} and \cite{adhikari20} (see \autoref{Sec:comparison_with_data}), where the mean redshift of the sample is $z=0.49$. When comparing with data radial distributions, we will present our results in units of $r/R_{200 \rm m}$. 

For the purposes of this paper, we define a galaxy to be an IllustrisTNG-defined subhalo \footnote{In SUBFIND a halo is generally called a subhalo, it does not necessarily mean it is within the virial radius of a larger structure.} with stellar mass greater than $10^8 h^{-1} M_{\odot}$. These are found using the SUBFIND algorithm \citep{Springel_2001, Dolag:2008ar}, each ``subhalo'', is a gravitationally bound structure that is defined using a friends-of-friends algorithm. The most bound particle is defined as the center of the halo. As we intend to study the total distribution of galaxies within the cluster and in its neighbourhood and make comparisons with data, we use all galaxies that are within 7$\cdot R_{200m}$ of the cluster halo center. %The center of the halo is defined as the position of the most gravitationally bound particle within the main halo, as determined by the SUBFIND algorithm. 

\section{Properties of Galaxies in Illustris}

\label{Sec:Properties}
In this section, we study the properties of galaxies in the IllustrisTNG simulations, focusing on their total and stellar masses, color, and star-formation rates. Notably, we find that galaxies around clusters in IllustrisTNG can be divided into three distinct groups in color-mass space. 
We examine the evolutionary history of these three groups in color-mass space along with their star-formation rates in order to probe the differences between them. We compare our color-mass splits to the color splits made in DES/ACT data \citep{adhikari20} and others in IllustrisTNG \citep{Nelson2018}.

\begin{figure}
    \includegraphics[width=8cm]{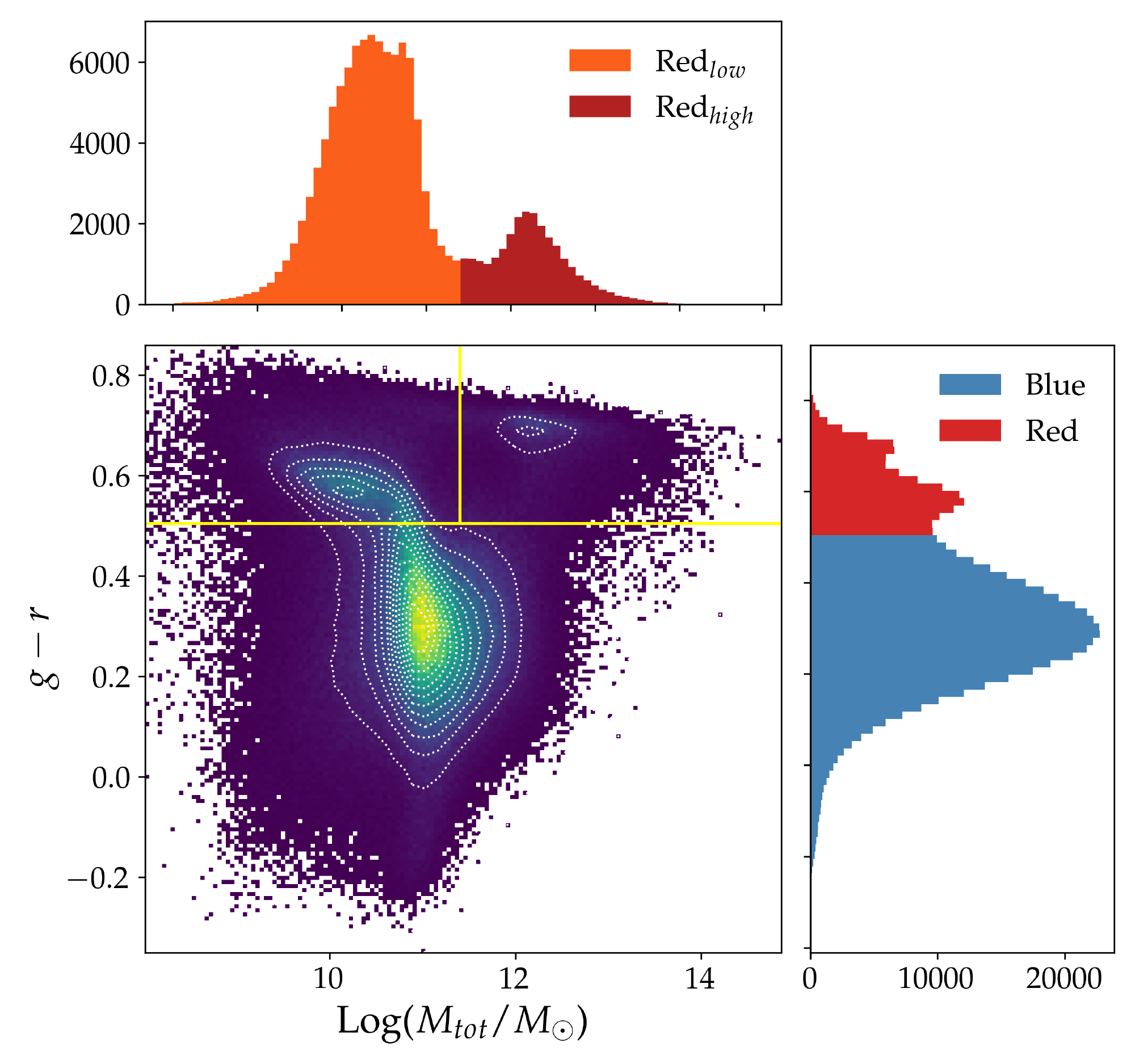}
    \centering 
    \caption{Color-mass plot of all galaxies at $z = 0.52$ in IllustrisTNG. Three distinct color-mass populations are visible. The one below $g-r$ of approximately 0.5 is "Blue", the one above $g-r = 0.5$ and with total mass less than $10^{11.4} h^{-1} M_{\odot}$ is "$\redl$", with the last group being "$\redh$". The top and right histograms display the one-dimensional distributions of the galaxies in color and total mass space. The top histogram of total mass includes only the galaxies defined as Red to illustrate  the mass cuts between the two Red populations. The color splits used at this redshift are indicated as yellow lines in the main plot.}
    \label{fig:colormass}
\end{figure}

\subsection{Color-Mass Populations}

\label{Sec:Color-mass}
We use the photometric data in IllustrisTNG for each galaxy to analyze the color distribution of galaxies in our sample. Photometric data in the $griz$ bands in Illistris are based on the SDSS Camera X Response Function (airmass = 1.3). We use $g-r$ for color throughout this paper. Note that the photometric information given in IllustrisTNG and used in this paper is not dust corrected and therefore our color sample is less "red" than would be expected in data.

In \autoref{fig:colormass}, we show the distribution of galaxies in the simulation at $z=0.52$ in color-mass space. The total mass of a galaxy in IllustrisTNG includes the stellar mass, dark matter mass, gas mass, and black hole mass associated with that given galaxy. We choose the total mass to define our populations as we focus on the dynamics of galaxies in the gravitational field of the cluster, which is determined primarily by their total mass, rather than their stellar mass. We note that there are three distinct populations in this space. While the usual red and blue sequence of galaxies is visible, we note that the galaxies in the red sequence can, in fact, further be divided into two distinct populations; we therefore define a different categorization of galaxies for this work: a Blue population with low $g-r$ colors, and two Red populations with low and high total masses. Note that the low and high mass Red galaxies are also, on average, lower and higher in $g-r$ color respectively.  In order to isolate these populations, we use a series of Gaussian mixture models and test the Bayesian information criterion (BIC) to obtain the most preferred models for splitting the color-mass space. The BIC for a Gaussian model with independent model errors is given by:
\begin{equation}
	\mathrm{BIC} = n \ln(\widehat{\sigma_e^2}) + k \ln(n) \
\end{equation}
where $n$ is the number of observations, or in this case the number of galaxies, $\widehat{\sigma_e^2}$ is the error variance, and $k$ is the number of parameters estimated, which increases with the number of Gaussian curves used to fit the data. A larger BIC value for the same data implies the model is less preferable. We first define a Red and Blue population using the distribution shown in the histogram on the right panel of Figure \ref{fig:colormass}. We find that a single Gaussian has a significantly higher BIC than using two Gaussians ($\Delta \textrm{BIC} \gg 10$). Although three Gaussians fit to $g-r$ data have a lower BIC than the two Gaussian fit, the resulting two Red populations from the three Gaussian fit are both bimodal in mass, thus there is no reason to prefer this split. We then define the exact boundary between the two resulting color-only populations by the intersection of the two Gaussian curves from the mixture model; at $z=0.52$, we split the populations at $g-r=0.505$. We then again use a Gaussian mixture model to determine the mass split between the two red populations in total mass, which we term $\redh$ and $\redl$. The BIC for the split using two Gaussians is again significantly lower than the single Gaussian, suggesting that splitting the two populations is preferred over treating them as a single "red" population. At $z=0.52$, the mass split divides $\redh$ and $\redl$ at a mass of $10^{11.4} h^{-1} M_\odot$. We repeat the same method of Gaussian mixture models to determine individual splits for each IllustrisTNG snapshot considered in our analysis. 

We note that our color selection is apparently a deviation from earlier studies of color evolution in IllustrisTNG like \cite{Nelson2018}, which studies the star-formation properties of central galaxies,  and \cite{Donnari_2020}, where the focus was mainly on quenched fractions using star-formation rate, and therefore was mainly studying a binary distribution of properties. As we will eventually study a population of galaxies, primarily around clusters, that are strongly biased objects, we isolate the clear overdensity of massive red galaxies living typically living in group mass halos. We therefore study the evolution of these three populations separately.

\subsection{The Star Forming Main Sequence}
\label{Sec:SFR}
\begin{figure}
    \includegraphics[width=9cm]{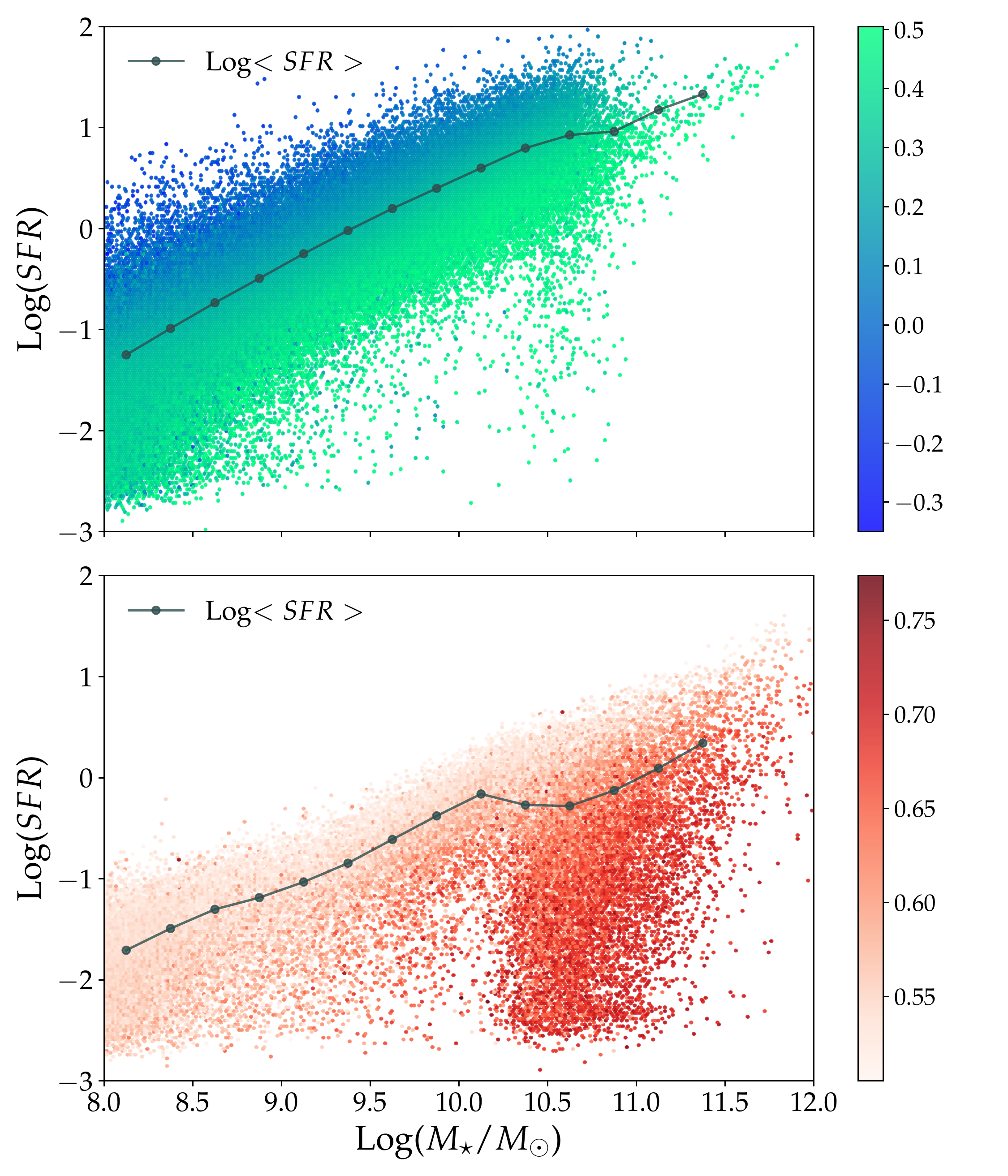}
    \centering 
    \caption{SFR main sequence for $z = 0.52$. These two plots show log SFR versus log stellar mass for all galaxies color-coded by their $g-r$ color, overplotted by a black line showing the log mean SFR in stellar mass bins. The upper plot displays the Blue sample, with the bluer and greener points indicating the higher and lower $g-r$ galaxies in the Blue population. Color is correlated with higher SFR at all stellar mass ranges for the Blue population, following the SFR main sequence. The lower plot includes the entire Red sample, i.e. both $\redl$ and $\redh$ galaxies, with redder points indicating redder galaxies. The reddest galaxies display a characteristic and exaggerated flattening of the main sequence.}
    \label{fig:SFR}
\end{figure}

We examine how the three galaxy populations defined in the color-$M_{\rm tot}$ space lie w.r.t to the Star-Forming Main Sequence (SFMS) \citep{Popesso_2018}. Galaxies tend to lie on this roughly linear main sequence, where the overall trend is that galaxies with larger stellar mass have higher log star-formation rates \citep{Donnari_2019}. In \autoref{fig:SFR}, we plot the logarithmically scaled star-formation rate and stellar mass, along with mean values in mass bins to represent the SFMS  \footnote{This plot does not include the galaxies to which IllustrisTNG assigns a value of SFR = 0 due to resolution effects, which corresponds to 2.5\% of Blue, 43\% of $\redh$, and 75\% of $\redl$ galaxies at $z=0.52$.}. The colors in the lower plot of the figure indicate $g-r$ color, with the lighter and darker red points largely coinciding with the $\redl$ and $\redh$ populations respectively.

We find that both Blue and $\redl$ galaxies tend to follow the SFMS and are consistent with the findings of \cite{Donnari_2019}. The mean star-formation rate for $\redl$ galaxies is lower than Blue ones. For both galaxy samples, we recover a  linear relationship between log SFR and log stellar mass for the mass range of $10^9$ to $10^{10.5} h^{-1}  M_{\odot}$. However, the high mass and higher $g-r $ $\redh$ galaxies, do not lie on the main sequence. As less red galaxies increase in log SFR along with mass, at high mass, we encounter galaxies with extremely low SFR that are significantly redder. This reflects what has been termed as a "flattening" or "bending" of the SFMS at stellar masses greater than $10^{10.5}$ $M_{\odot}$, which is most similar to our $\redh$ sample \citep{Popesso_2018}.

In the following subsection we examine evolutionary processes associated with these galaxies.

 %This population with depressed SFR is also visible in the main sequence plots of \cite{Donnari_2019} but appears less distinguishable due to the relative scarcity of $\redh$ galaxies as compared to Blue and $\redl$ when overplotted in the figures. We concur with \cite{Donnari_2019} that a bimodality in log SFR is much less obvious than the color bimodality, however, we emphasize that in two-dimensional log SFR and stellar mass space there is nonetheless a clear bimodality visible when isolating Red galaxies.

\label{Sec:pophist} 
\subsection{Galaxy Population Histories in Color-Mass Space}

\begin{figure*}
    \centering
    \includegraphics[width = 18cm]{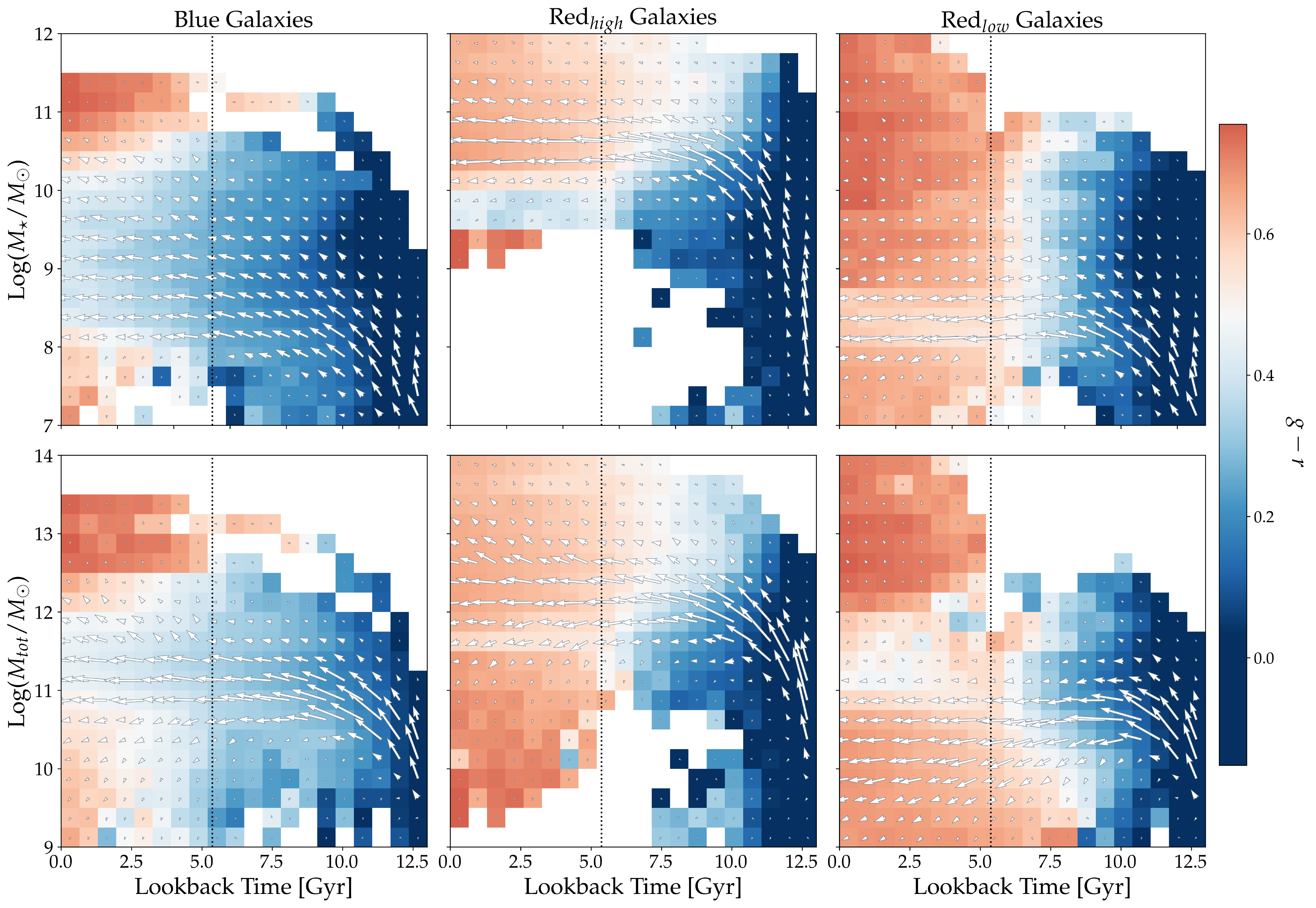}
    \caption{
    % Evolution of a sample of 3000 galaxies selected at $z=0.52$, sampled equally from the three galaxy populations, (Blue, $\redh$, and $\redl$). The pixel color in each panel shows the average lookback time of all the galaxies in that pixel; for example, the pixel color in the top-left panel, which only includes galaxies considered Blue at $z=0.52$ (lookback time of 5.3 Gyrs), is colored in each bin by the average lookback time of the galaxies in that bin. The arrows represent the average motion in color-mass space between consecutive simulation snapshots of the galaxies located within a given color-mass bin at a given snapshot, such that each step in the motion of the galaxy in color-mass space is reflected in the arrows. The length of the arrows reflect the magnitude of the motion of galaxies in a given bin to their subsequent bin in the next snapshot, while the arrow thickness represents the number of galaxies within each bin.
    Evolution of a sample of 3000 galaxies sampled equally from the three galaxy populations: Blue (left), $\redh$ (middle), and $\redl$ (right). The pixel color in each panel shows the average $g-r$ color of all the galaxies in that pixel. 
    %; for example, the pixel color in the top-left panel, which only includes galaxies considered Blue at $z=0.52$, is colored in each bin by the average $g-r$ color of the galaxies in that bin. 
    The length and direction of the arrows represent the average change in mass and time  between consecutive simulation snapshots.
    %of the galaxies located within a given mass-time bin at a given snapshot, such that each step in the motion of the galaxy in mass-time space is reflected in the arrows. 
    %The length of the arrows reflect the magnitude of the motion of galaxies in a given bin to their subsequent bin in the next snapshot, while 
    The arrow thickness represents the number of galaxies within each bin. The vertical black dotted line in each panel indicates the  lookback time at $z = 0.52$, the redshift at which the populations are identified in this figure.
    }
    \label{fig:colorev}
\end{figure*}

We use the LHaloTree merger tree for the IllustrisTNG 300-1 simulation to examine the histories of the three color-mass populations defined in the previous section over a redshift range of $z = 14$ to $z = 0$. At $z=0.52$, we randomly sample 1000 each of Blue, $\redh$, and $\redl$ galaxies from IllustrisTNG, tracing their histories in mass and color both forwards and backwards in time. We use the main progenitor branch and descendent branch to trace mergers and histories. The main progenitor of a subhalo is defined as the parent subhalo which has the largest combined mass in its history. In \autoref{fig:colorev}, we plot the pathways which each of our three populations take in mass-time space, showing the average color of galaxies at that time and mass. We highlight some of the most notable features below.

Firstly, as expected, we find that all galaxy populations begin as blue galaxies with low stellar masses at early times. With time, galaxies evolve into redder colors, but the color evolution for the three populations are distinct from each other. In general galaxies become red either when they reach a certain mass, or by environmental effects that is also accompanied by a loss in the total mass of the halo presumably due to tidal effects. We also note that the mass assembly itself is different, in particular galaxies that are in the $\redh{}$ population tend to assemble their mass earlier in time than lower mass Blue and $\redl{}$ galaxies. 
%Firstly, we find that all galaxies start out in the bottom right as low mass (stellar or total) and blue and eventually migrate to the top right corner, whether by directly becoming $\redh$ galaxies, or becoming $\redl$ galaxies and eventually merging into larger galaxies. Each pixel is colored by the average $g-r$ color of all galaxies in that location in mass-time space.

%Note that the pixel-color evolution across the diagonal is different for Blue and Red galaxies. Red galaxies live in halos that typically assemble their masses at early times, while that is not the case for Blue galaxies \sa{(cite relevant papers)}.

%\mb{Not all of them end up there \textit{yet} - we should specify that red lows eventually merge into red highs to go into the top right corner}

The $\redh$ population at $z = 0.52$ originates as Blue galaxies, which first increase in total and stellar mass without significantly reddening, and then upon reaching a critical mass, quickly quench into the massive $\redh$ population. This is indicative of an AGN driven feedback that is triggered at a critical mass threshold  \citep{Donnari_2020, Terrazas_2020, Nelson2018, Weinberger_2018} and progresses relatively quickly. At later times, some of these $\redh$ galaxies lose total mass while maintaining their stellar mass, presumably from tidal mass loss by merging into cluster-sized objects. $\redl$ galaxies, on the other hand, begin to quench at a lower stellar and total mass, and quickly enter a phase where they begin to lose total mass, implying that their halos are getting stripped as they undergo infall into a larger object. We note that the mass loss begins after the galaxies are significantly reddened. We also note that while a part of the $\redl$ galaxies at $z = 0.52$ appears to become a population of high $g-r$ and high mass galaxies at late times (presumably through mergers and the like, the arrow size in those bins indicates that it is a relatively small fraction of the population compared to those that maintain or lose mass as the redden further. 

We calculate the average time it takes the galaxies in our $\redl$ and $\redh$ samples selected at $z = 0.52$ to travel from a $g - r$ value of 0.0 to 0.5. Galaxies that end up as $\redl$ take on average $4.79 \pm 1.33$ Gyr to traverse that color range. Galaxies that end up as $\redh$ tend to quench faster with an average timescale of $4.08 \pm 1.66$ Gyr. These distinct Red populations are the products of different mechanisms with different rates. The large spread of $\redh$ quenching times could also indicate that the population is formed through multiple quenching mechanisms with at least one acting rapidly compared to the quenching of $\redl$ galaxies.  

% \td{this paragraph needs to be edited, deleted, or reconsidered after the plot change}Finally, we note that there is bifurcation in the Blue population along the diagonal. At early times, a fraction of the galaxies tend to get bluer as they increase in mass, indicating an early increase in star-formation, while others tend to get quenched. 
We note that, at late times, Blue galaxies at $z=0.52$ that end up quenched by $z=0$ do not appear to lose total mass, which is likely a result of them mostly being in fields and voids (low density regions).

Overall, we find evidence for two dominant quenching scenarios, consistent with those proposed in literature. Mass-triggered quenching and environmental quenching are visible and distinguishable in these two distinct Red galaxy populations in IllustrisTNG. At our given fiducial redshift, the low-mass Red galaxies ($\redl$) are typically environmentally quenched, reflecting the effects of environmental processes such as tidal stripping, ram pressure stripping, starvation, and harassment \citep{GunnGott72, Abadi:1999qy, Contini_2020, Donnari_2020}. On the other hand, the massive red galaxies ($\redh$) are quenched by both internal processes, such as AGN feedback, and mergers.

\begin{figure*}
    \includegraphics[width=18cm]{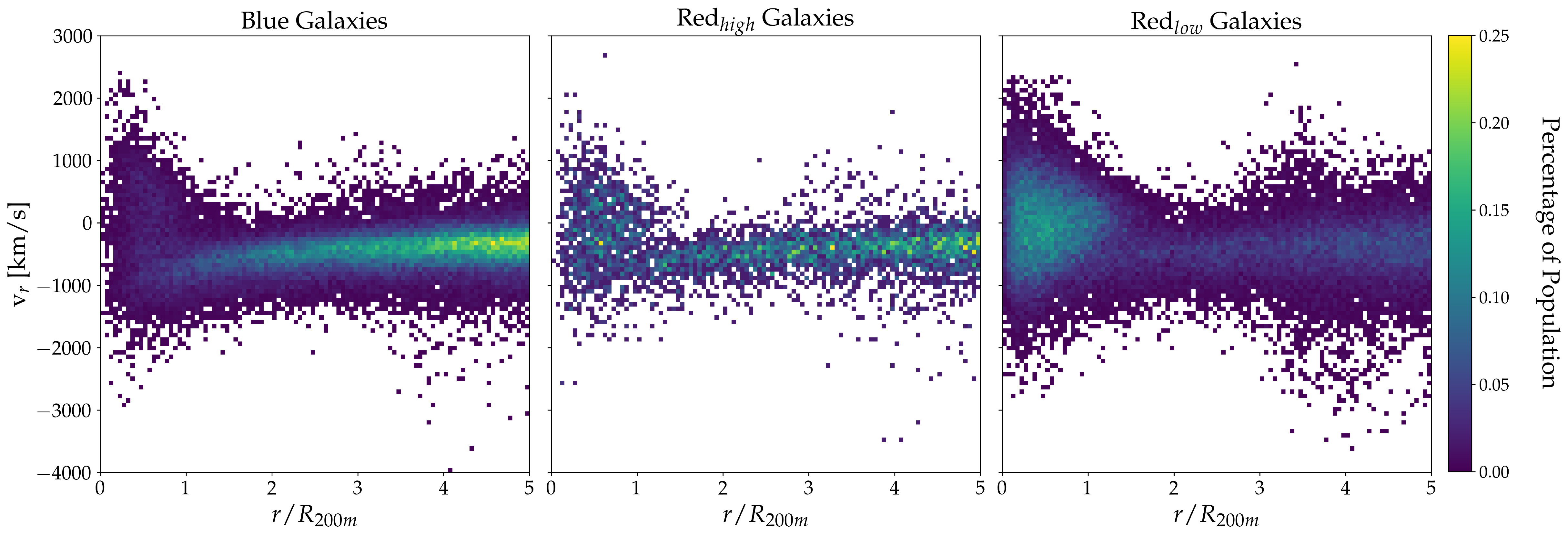}
    \centering 
    \caption{Stacked 3d phase space diagrams for the three color populations (Blue, $\redl$, and $\redh$ going from left to right) at $z = 0.52$. The color indicates the percentage of the population in that 2d bin, i.e yellower bins indicates the presence of more galaxies. All three populations show clearly defined infall streams and multistreaming regions. Blue galaxies are predominantly located in the infall stream, with their density falling off towards the multistreaming region. $\redh$ galaxies (middle panel) are split evenly between infall and multistreaming regions, with the lowest radial velocity dispersion in the infall stream. $\redl$ galaxies are present across the entire virialized region as well as the infall stream, and have the highest radial velocity dispersion in the infall stream.}
    \label{fig:phase_space_colors}  
\end{figure*}

\section{Phase Space Distribution of Galaxy Properties}
\label{Sec:PhaseSpace}

In this section we study the distribution of galaxies and galaxy properties in the phase space of dark matter halos. The phase space of halos is the two-dimensional space of the radial velocity and the radial distance from the halo center. This space provides an intuitive way to look at the evolution of a dark matter halo as different regions of this space can be mapped on to different times in the halo's formation history. In the idealized cold dark matter spherical evolution \citep{1984ApJ...281....1F}, every particle follows a unique trajectory in this space as the dark matter sheet warps with time. Galaxies that are collisionless tracers of the dark matter potential are also expected to follow similar phase space trajectories.
% \td{unsure if this previous sentence should have "that" or ",which", it changes the meaning.} 

Phase space diagrams of the radial velocities of particles prominently display two distinct regions in and around clusters: the infall stream and the multistreaming region. Defined as the steepening of the density profile at the location of first turnaround, the splashback radius traces the boundary between these two regions visible in phase space. Matter on its first infall resides in the infall stream, extending across the range of radii in the diagram.  Matter already "virialized" and orbiting in the cluster resides in the multistreaming region, extending to approximately the splashback radius with a range of radial velocities. 

Moreover, the spherically averaged radial density profile of a halo is the integral of the phase space distribution in the velocity direction.
%\ts{[}, i.e. the number of galaxies in a given radial location is the sum of the number of velocity streams at that location;\ts{] Do we need this explanation? The sentence proceeding it is already very intuitive and easy to understand.}
Therefore, galaxies occupying different regions in phase space imprint their signatures on the observed shape of the density profile. While this property was exploited in \cite{adhikari20} to infer the times of infall of different galaxy populations and also to constrain a simple quenching model, we directly test this hypothesis in this work using hydrodynamic simulations to explore whether observable properties of galaxies indeed separate out in phase space and what their distributions help us infer about galaxy evolution.

\subsection{Distribution of Galaxy Color}
In this section, we study the distribution of galaxy color in the phase space of dark matter halos. We divide our galaxies into three populations namely Blue, $\redh$ and $\redl$, as described above, and analyze their distribution in phase space. In \autoref{fig:phase_space_colors}, we plot the stacked phase space for the three color-mass populations at z = 0.52. We divide the two-dimensional space of v$_r$ and $r/R_{200m}$ into 100 bins along the radial direction and 100 bins along the velocity direction. The color of the bins indicates the number density of galaxies in each bin. We note that the three different populations display distinct behavior in phase space. We highlight the most significant features of these distributions: Blue galaxies are dominant in the infall stream of the cluster whereas the low-mass Red galaxies ($\redl$) occupy the full extent of the virialized region along with the infall stream. The high mass Red galaxies ($\redh$) appear to be concentrated near the center of the cluster and appear to have a lower radial velocity envelope compared to both the Blue and $\redl$ population. We also note that in the infall regions(~$r/R_{200\rm m}>1.5$ Mpc $h^{-1}$), i.e., beyond the splashback radius, both Blue and $\redl$ galaxies tend to have a broad radial velocity dispersion compared to the $\redh$ galaxies.

\begin{figure*}
    \includegraphics[width=18cm]{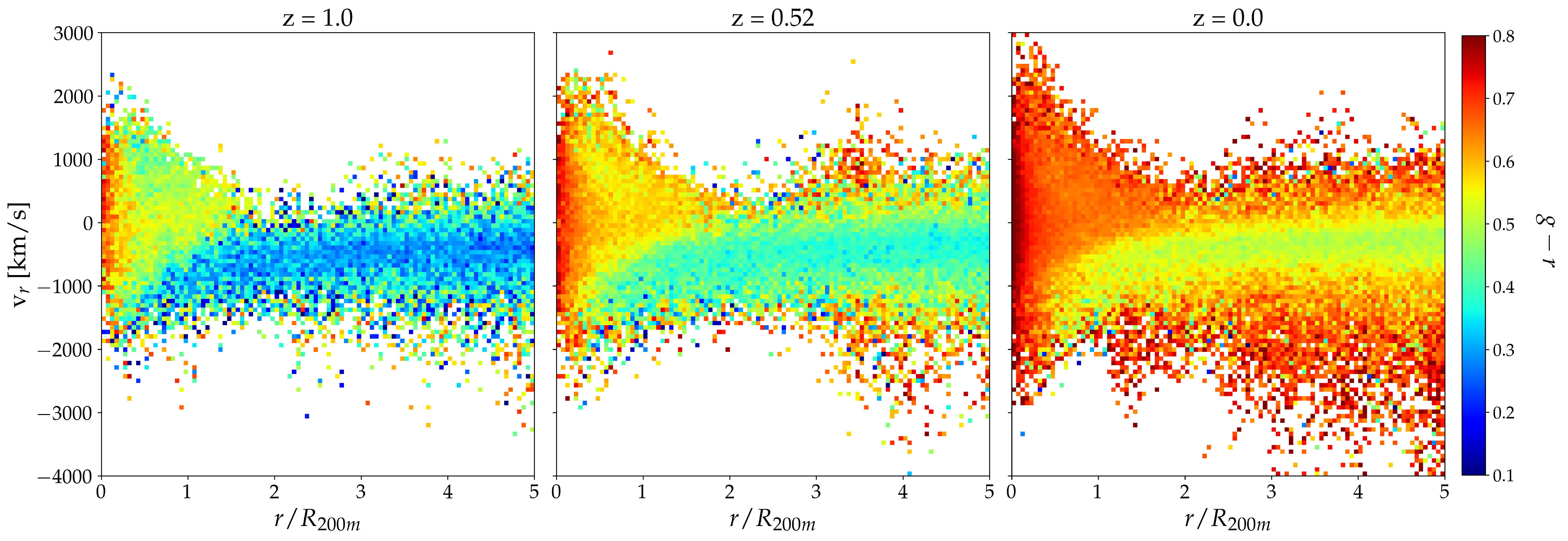}
    \centering 
    \caption{Stacked 3d phase space diagrams for all galaxies in our sample at three redshifts as indicated above each panel. The points are colored by the mean $g - r$ color in a given bin of the 2d histogram. %Note that the density of galaxies in each histogram bin is not represented in this plot. 
    At earlier times, the infall stream is predominantly composed of blue galaxies, with only a few redder galaxies at high velocities. At later times, as more structure grows, the infall stream widens significantly, with higher (absolute value) velocities dominated by redder galaxies. The red galaxies at high radial velocities are undergoing infall as parts of groups.}
  	\label{fig:phase_space_3z_weighted}
\end{figure*}

The phase space distribution of these different populations can be well understood by their evolutionary pathways. Firstly, if Blue galaxies are preferentially quenched inside the cluster, they would be most dominant in the infall stream. Also, if Blue galaxies get quenched into Red ones, the multistreaming region in the phase space would be dominated by Red galaxies with time, which is clearly seen in the rightmost panel of \autoref{fig:phase_space_colors}. However, we note that the most massive Red galaxies presumably fall into the cluster as central galaxies of groups that are red even before infall into the cluster. These massive objects tend to sink to the center because they lose momentum due to dynamical friction and self-friction (see \autoref{Sec:DynFricSec}), and appear to be confined within a low radial velocity envelope, that only extends out to about $\sim 0.5 R_{sp}$.

Analyzing the infall streams, i.e. with $r>R_{sp}$, we note that the $\redl$ and Blue galaxies tend to extend out as clouds from the narrow infall stream. This is due to the fact that a significant fraction of these galaxies fall in as satellites of higher mass groups, with their own phase space structures. We note that $\redl$ galaxies tend to live in these high velocity clouds. This further distinguishes them from $\redh$ galaxies which tend to live in the narrow infall stream, likely as field galaxies or as centrals of the aforementioned infalling groups. In \autoref{app:los}, we also plot the line of sight velocity dispersions of the three populations in three cluster-centric radial bins. The line of sight velocities are often used in spectroscopic surveys to infer quenching timescales \citep{2016MNRAS.463.3083O, Adhikari:2018nxy} and to infer halo masses \citep{Anbajagane:2021gfx}, we note that the dispersions can change significantly as a function of radius and galaxy color. 

In \autoref{fig:phase_space_3z_weighted}, we plot the stacked phase space of the clusters with bin color weighted by the mean $g-r$ color of all the galaxies in that bin. We plot this color-weighted phase space at three different redshifts, $z= 0, 0.52, 1.0$. We note that at all redshifts, the infall stream remains dominated by bluer galaxies. Generally, the galaxy population around a cluster gets redder with time. We note particularly that the velocity dispersion of the  infall stream gets larger at late times as structure grows, and the higher velocity tails are occupied by red galaxies quenched pre-infall.

\begin{figure}
     \includegraphics[width=8.5cm]{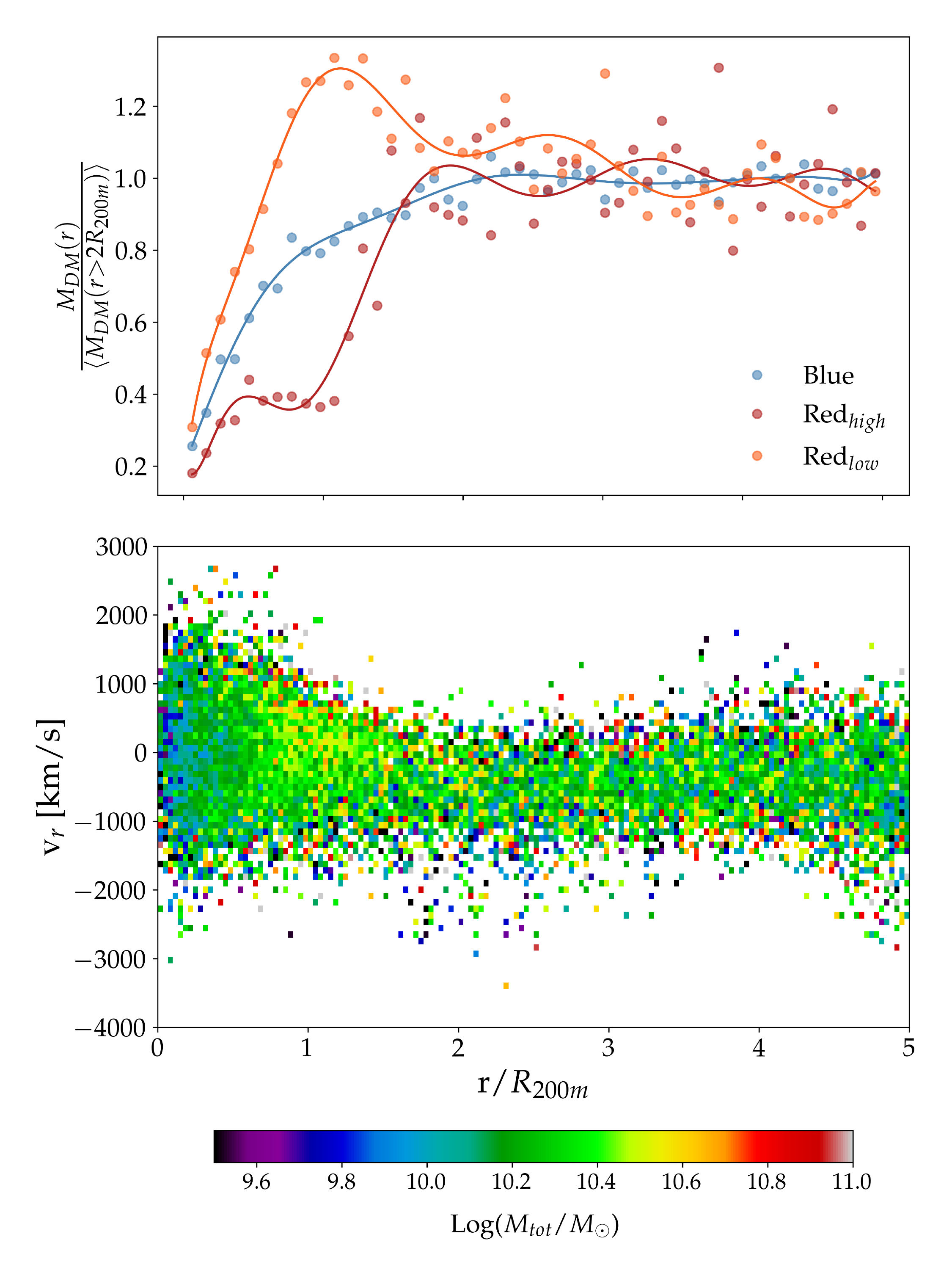}
    \centering 
    \caption{\textit {Top}: The points indicate the ratio of the median dark matter mass of galaxies in radial bins to the median dark matter mass outside the cluster, between $2-7 \cdot R_{200m}$. The ratio is plotted for each of the three color-mass populations at $z = 0.52$. %Each point on the diagram is located at the center of a radial bin, for which we determine the ratio of the median dark matter mass within that bin to the median dark matter mass in cluster exteriors (outside $2 R_{200m}$). 
    The solid lines show a polynomial fit to the  points. %We do an 11th order polynomial fit to demonstrate the change in this ratio. 
    Blue and $\redh{}$ galaxies have strictly decreasing dark matter masses as they fall into cluster centers. The $\redl$ population displays a higher dark matter mass ratio between between pericenter and splashback, before decreasing towards the center. This is potentially due to the quenching of infalling Blues into Red. %$\redh$ experience a more rapid decrease in dark matter mass ratio due to a similar Blue quenching effect in addition to dark matter stripping. 
    \textit{Bottom}: $\redl$ galaxies in phase space, weighted by the mean total mass of the galaxies in each bin. A triangular region of distinctly higher total mass is visible near $R_{200m}$, likely a result of the same effect causing the dark matter fraction to  increase at the same radii.}
    \label{fig:Mdm_fraction_curve}
\end{figure}

In the following section we examine the different components of these galaxies in the phase space, i.e. we study the relative ratios of gas, stellar mass and dark matter as a function of their radial velocity and distance from the cluster center.

\subsection{Distribution of Dark Matter and Baryonic Mass}
\label{Sec:matterdist}
In this section, we study how the relative fractions of stars, dark matter, and gas of a galaxy change as a function of its location around the halo in phase space. In particular, we look at the dark matter and gas mass ratios for the three different galaxy populations defined previously. Primarily, inside the halo, we expect there to be two effects of significance that affect the galaxy star-formation property. The removal of gas due to ram-pressure, heating, etc. and the general tidal disruption of the extended dark matter halo leading to a reduction in dark matter mass. 

Firstly, in Fig. \ref{fig:Mdm_fraction_curve}, we see how the dark matter mass ratio changes for the three different populations as we move closer to the center of the host cluster halo. The curves in the figure show the ratio of the dark matter mass at a given radius to the median dark matter mass between 2-7$\cdot R_{200m}$. We note that while the massive Red galaxies and the Blue galaxies show a steady fall-off of dark matter mass ratio as we move towards the center, as is expected from tidal disruption of halos, the $\redl$ galaxies appear to show an increase roughly near $\sim R_{200m}$, before the mean mass begins to drop. To investigate the origin of this effect, we look at the distribution of the mean total mass of galaxies at different locations in phase space. In the lower plot of \autoref{fig:Mdm_fraction_curve}, the $\redl$ galaxies between pericenter and splashback appear to have higher total mass than elsewhere in the sample. A possible explanation for this effect is the quenching of Blue galaxies into low-mass Red galaxies between first pericenter and first apocenter (splashback); We divide our galaxies into populations based on their present day color and total masses, and Blue galaxies tend to have higher total mass than low-mass Red galaxies (see Fig.\ref{fig:colormass}). Therefore, a recently quenched galaxy, if on its first orbit with relatively low tidal disruption, might appear red but still have a significantly high total mass. We note that the radial distribution of the total mass, that can be probed by observations like weak lensing of satellite galaxies, carries information about the evolution of galaxy quenching.

\begin{figure*} 
    \includegraphics[width=\textwidth]{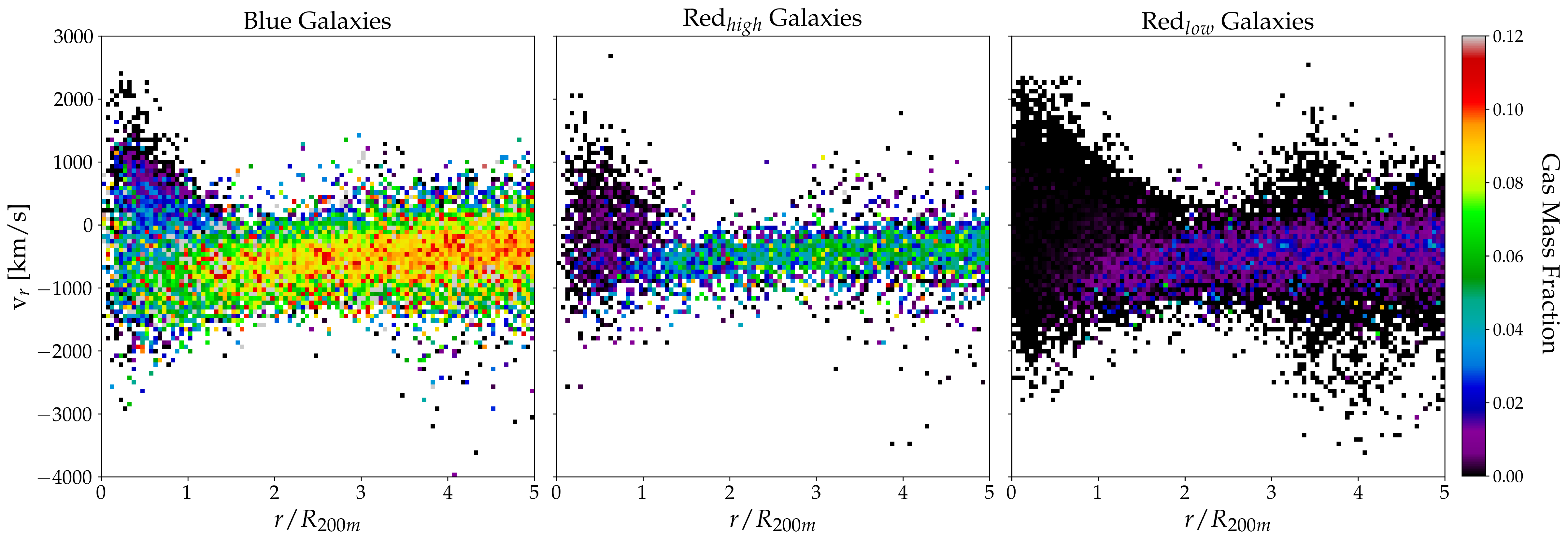}
    \centering 
    \caption{Stacked phase space diagrams for the three color populations (Blue, $\redl$, and $\redh$ going from left to right) at $z=0.52$, with each bin in the 2d histogram weighted by the mean ratio of gas mass to total mass of the galaxies in that bin. Blue galaxies have the highest proportions of gas, with higher velocity galaxies having less gas. The effects of gas stripping in cluster interiors are clearly visible as blue galaxies in the multi-streaming region lose nearly all of their gas. $\redh$ galaxies demonstrate a similar behavior to the Blue galaxies. $\redl$ galaxies also lose what little gas they have as they fall in. Note that $\redl$ galaxies with the high radial velocities in the infall stream show a level of gas stripping comparable to those in the cluster interior, suggesting they were pre-processed as parts of smaller clusters.}
    \label{fig:phase_space_gas}
\end{figure*}

In \autoref{fig:phase_space_gas}, we plot the mean gas mass fraction of galaxies at every location in phase space. The three different populations show significantly different behavior that is reflective of their evolutionary history in the halo. The Blue galaxies, that are star-forming, naturally tend to have the highest gas mass fractions among the different populations. In the infall stream these galaxies have nearly $10 \%$ of their total mass made of gas. As we move within the halo, we note that post-pericenter the gas mass fractions appear diminished. This can be explained by the gas being severely stripped once the galaxies fall into the cluster host, due to effects like ram-pressure stripping, which should strongly depend on orbital history and leave a signature on the phase space. Similarly, we note that the $\redl$ galaxies inside the cluster multistreaming region seem to have lost almost all of their gas. We also note that, although galaxies in the infall stream overall have higher gas mass fractions, the $\redl$ galaxies that lie in the extended clouds tend to have gas mass fractions comparable to those that are within the clusters. This shows that a fraction of the galaxies that fall in as a part of groups have already lost significant amounts of their gaseous halos. The $\redh$ galaxies, which we expect to be centrals of groups, tend to have gas fractions significantly higher than $\redl$ galaxies but lower than Blue galaxies in the infall stream. $\redh$ galaxies show a general decrease of gas fraction within the cluster similar to the other two population. We note that the $\redh$ in the multistreaming region seem to show lower gas fractions for galaxies in the positive velocity stream as they lose gas after pericenter passage. explanation. In Appendix \ref{app:phasespace} we also show how other properties, such as the dark matter mass fraction and the black hole mass fractions, behave in phase space. 

The various evolutionary histories of galaxy star-formation leave distinct signatures on the phase space of dark matter halos. As the radial number density profile of galaxies is simply an integration of the phase space along the radial velocity direction, these distinct pathways will leave specific signatures on the density profiles of galaxies and their shapes. In the next section, we analyze how the various properties of galaxies affect the profile shapes and their slopes.

\section{Radial Distribution of Galaxies and Splashback radius}
\label{Sec:Profiles}

%\subsection{Measurement of Galaxy Number Density Profiles} 
At a given redshift, we use the following procedure to produce a single galaxy number density profile for a given population of galaxies. For each halo, we compute the spherically averaged galaxy number density profiles in 15 logarithmically spaced radial bins. The bins extend to $7\cdot R_{200m}$ for each individual cluster, as opposed to a fixed physical distance as in other works. This allows for a higher signal-to-noise result when stacking the profiles, as the splashback radius scales with cluster properties like $R_{200m}$. The individual cluster profiles are stacked to produce a single average profile. The covariance matrix and error bars on the profiles are calculated using the Jackknife resampling method \citep{Norberg09}.

\subsection{Model Fitting of Galaxy Number Density Profiles} 
We model the galaxy number density profiles using the fitting formula described in \citep{Diemer:2014gba}. The 3d galaxy number density profile is defined by a virialized Einasto component and an outer infall component with some transition region \citep{Einasto1965}. This is expressed as:

\begin{equation}
    \rho(r) = \rho_{\text{inner}}(r)f_{\text{trans}}(r) + \rho_{\text{outer}}(r)
\end{equation}

These terms are given by the following equations:
\begin{equation}
    \rho_{\text{inner}} = \rho_s \exp{\left(-\frac{2}{\alpha}\left(\left(\frac{r}{r_s}\right)^\alpha-1\right)\right)}
\end{equation}
\begin{equation}
    f_{\text{trans}}(r) = \left[1+\left(\frac{r}{r_t}\right)^\beta\right]^{-\gamma/\beta}
\end{equation}
\begin{equation}
    \rho_{\text{outer}}(r) = \rho_0 \left(\frac{r}{r_0}\right)^{-s_e}
\end{equation}

This is the model that is also used in the measurement of splashback radius from projected galaxy number density profiles \citep{More:2016vgs, Baxter:2017csy, Chang:2017hjt, Shin19, adhikari20}. However, unlike the data model, we do not include a term for cluster miscentering. Our model therefore has a total of eight parameters from the halo model \citep{Diemer:2014gba}. The parameters $r_0$ and $\rho_0$ are degenerate with one another, thus we fix $r_0 = 1.5 \text{ Mpc } h^{-1}$. We fit the 3d galaxy number density profile with the eight-parameter model above using a Markov Chain Monte Carlo (MCMC) method implemented in the \textsc{emcee} package, along with a jackknife estimate of the covariance matrix, adopting a Gaussian likelihood. The priors for each parameter are specified in \autoref{tab:my_label}.
Since the uncertainty on the density profiles from the simulation is relatively small, we only show the best-fit curves in the subsequent sections.
\begin{table}[h!]
    \small
    \centering
    \begin{tabular}{c|c}
         Parameter & Prior \\
         \hline
         $\log(\frac{\rho_s}{h^3 \textrm{Mpc}^{-3}})$ & [-8.0, 8.0] \\
         $\log(\frac{\rho_0}{h^3 \textrm{Mpc}^{-3}})$ & [-8.0, 8.0] \\
         $\log(r_s [h^{-1} \textrm{Mpc}])$ & [$\log(0.1), \log(5.0)$] \\
         $\log(r_t [h^{-1}\textrm{Mpc}])$ & [$\log(0.1), \log(5.0)$] \\
         $\log(\alpha)$ & $\mathcal{N}(\log(0.22), 0.6^2)$ \\
         $\log(\beta)$ & $\mathcal{N}(\log(6.0), 0.2^2)$ \\
         $\log(\gamma)$ & $\mathcal{N}(\log(4.0), 0.2^2)$ \\
         $s_e$ & [0.1, 10] \\
        %  Parameter & Prior \\
        %  \hline
        %  $\log(\rho_s)$ & [-8.0, 8.0] \\
        %  $\log(\rho_0)$ & [-8.0, 8.0] \\
        %  $\log(r_s)$ & [$\log(0.1), \log(5.0)$] \\
        %  $\log(r_t)$ & [$\log(0.1), \log(5.0)$] \\
        %  $\log(\alpha)$ & $\mathcal{N}(\log(0.22), 0.6^2)$ \\
        %  $\log(\beta)$ & $\mathcal{N}(\log(6.0), 0.2^2)$ \\
        %  $\log(\gamma)$ & $\mathcal{N}(\log(4.0), 0.2^2)$ \\
        %  $s_e$ & [0.1, 10] \\
    \end{tabular}
    \caption{Prior ranges for the eight model parameters. $\mathcal{N}(x, \sigma^2)$ is a Gaussian prior with mean $x$, standard deviation $\sigma$. We do not include a miscentering parameter.}
    \label{tab:my_label}
\end{table}

\begin{figure*}[t!]
    \includegraphics[width=18cm]{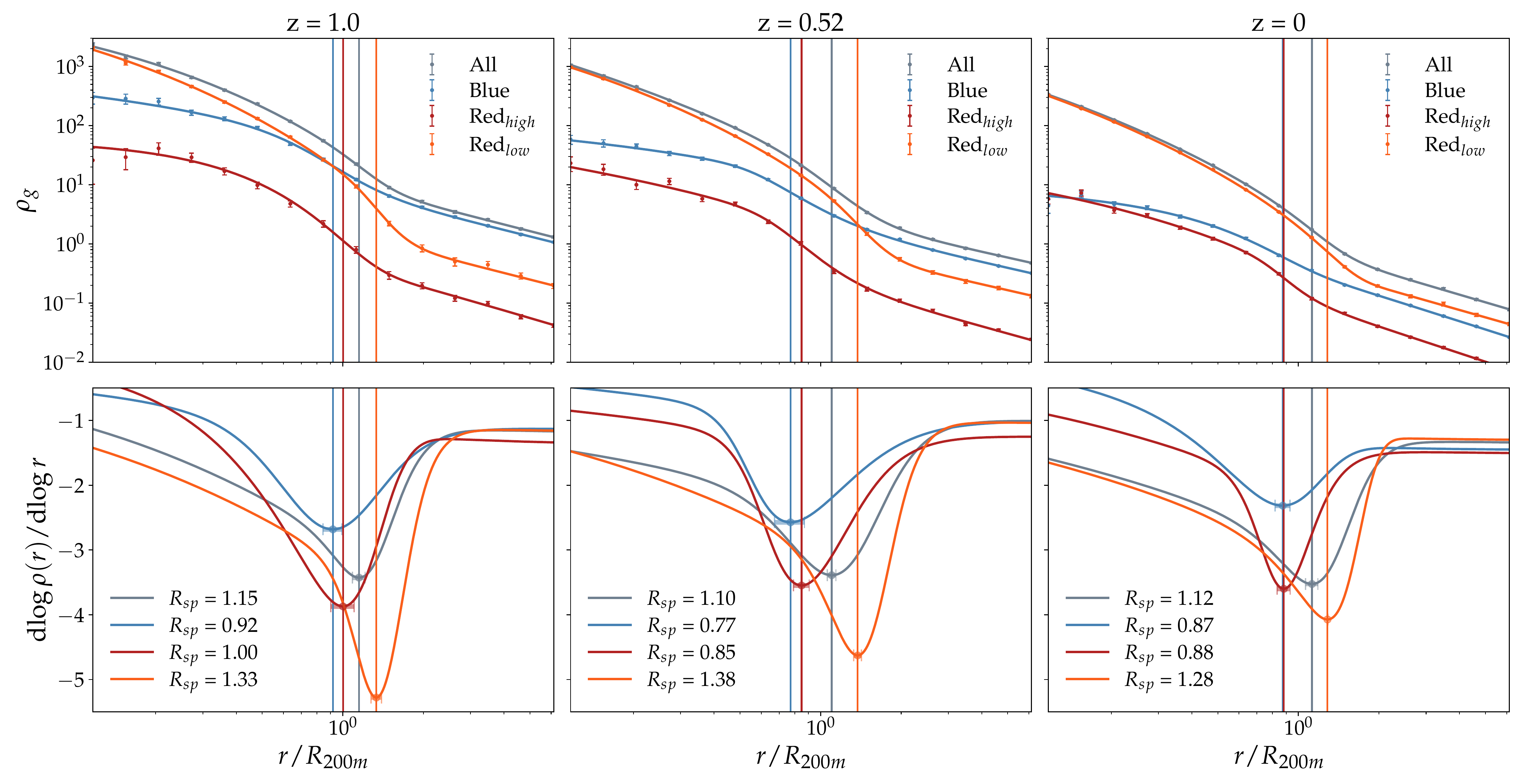}
    \caption{Galaxy number density profiles at $z=1.0, 0.52, 0$ (left to right) for each  of the three different color-mass populations (upper panels), with their corresponding fitted logarithmic slope profiles (lower panels). The splashback radius, defined as the minimum of the slope, is shown by the vertical line. The profiles are given for each of the three color split populations as well as the total profile, as indicated in the legend in the upper panels.}
    \label{fig:profiles}
\end{figure*}

% We identify the splashback radius as the location of the minimum logarithmic slope and estimate its error using the standard deviation of the MCMC samples... \mb{what should go here?}

%In this section, we examine the splashback feature and the number density profiles in the sample of halos we selected from the IllustrisTNG 300-1 simulation. We see how the radial distributions of the three populations behave and study the evolution of their splashback radius as a function of redshift. We also compare our results from a full hydrodynamic simulations to the observed distribution of colors and their splashback radius from \cite{Shin19} and \cite{adhikari20}. A key conclusion was that the existence of a splashback feature for the Green galaxies (the population with $g-r$ values in between Red and Blue) indicates that complete quenching takes longer than one pericentric passage through the cluster. Moreover it was found that the color populations have distinct profile shapes as well as splashback radii and these could be related to times since infall and quenching timescales. We perform a similar analysis for our sample and color-mass selection, and expand on the results from \cite{Shin19} and \cite{adhikari20} by examining redshift evolution of splashback.

\subsection{Color Population Profiles}

We evaluate the galaxy number density profile around the cluster sample for the three color populations in the redshift range $1 > z > 0$. The profiles and their corresponding slopes for three selected redshifts ($z = $ 0, 0.52, 1.0)  are shown in  Fig. \ref{fig:profiles}. To reiterate, these are stacked profiles for halos with mass greater than $ 5\times 10^{13} M_\odot h^{-1}$. In the top panels, the measured density profiles and the best-fit models using the DK14 profile are shown.
%In the top panel the points are the measured profiles and the lines are the best-fit models obtained by fitting the DK14 profile. 
The bottom panels show the logarithmic slope of the fitted profiles. \footnote{Note that the profiles decrease in amplitude with redshift. This is due to the fact that $\rho_g$ is calculated in bins defined by $R_{200m}$ which increases at lower redshift.}

The essential thing to note is that, at any given redshift, the location of the minimum slope of the density profile is sensitive to the properties of the galaxies and encode evolutionary information. The three different samples in our study, we expect, arise from three distinct evolutionary histories. Our work demonstrates that these histories leave characteristic signatures in their radial distributions around the clusters.
%in the cluster mass. 

\textit{Red galaxies:} Firstly, we note that the splashback radius of the $\redl$ galaxies extends farthest out in units of $r/R_{200m}$ at all redshifts. At $z = 0.52$, for example, the splashback radius of $\redl$ galaxies is 25\% higher than the total profile. We expect these galaxies to trace the full extent of the dark matter multi-streaming region of the phase space. Note that Fig.\ref{fig:phase_space_3z_weighted} shows that galaxies, after pericentric passage tend to get subsequently less blue with time, therefore the galaxies that do survive to the phase space boundary, or splashback radius, are preferentially red (as blue galaxies also transform to red). Moreover, while the most radial orbits reach the farthest distances at splashback \citep{Diemer:2017ecy,Banerjee:2019bjp}, they are also likely to be quenched faster than less radially orbiting galaxies, due to the smaller pericenters where they experience higher pressure drag.

The $\redh$ population shows significantly different behaviour from their low-mass counterparts. The location of the minimum slope for this population is shifted to a smaller radii with respect to the full population. For $z = 0.52$, the splashback radius of $\redh$ is 22\% lower than the All-galaxy profile. This can be understood as these massive galaxies are drawn towards the center due to dynamical friction. The location of their minimum in slope in fact corresponds to the ``splashback'' radius of that population, i.e. the apocenter itself is drawn to a smaller radius.

\textit{Blue galaxies:} We find that the Blue galaxies show a shallower splashback feature as compared to either of the Red galaxy populations, however, we do not see similar results to blue galaxies from \citep{Shin_2021} or \citep{adhikari20}. 
%\mb{that was from older results, should've bbeen removed} \sa{is your error bar very different from redh or redl :)?}
For $z = 0.52$, the splashback radius of Blue galaxies is 30\% lower than that of the All-galaxy profile. Blue galaxies have a very shallow inner profile, that becomes shallower at later times\ts{:} at $z=0$, the slope becomes zero (flat density) deep inside the cluster. The location of the minimum slope is significantly lower than that of the $\redl$ galaxy profile and the All-galaxy profile, but is relatively close to that of the $\redh$ population. In \autoref{Sec:comparison_with_data}, we discuss how the behavior of the blue galaxies compares to data and discuss possible reasons for the shallower slope and prominent slope minimum of the Blue galaxy profile in IllustrisTNG.

\begin{figure}
    \centering
    \includegraphics[width=8.5cm]{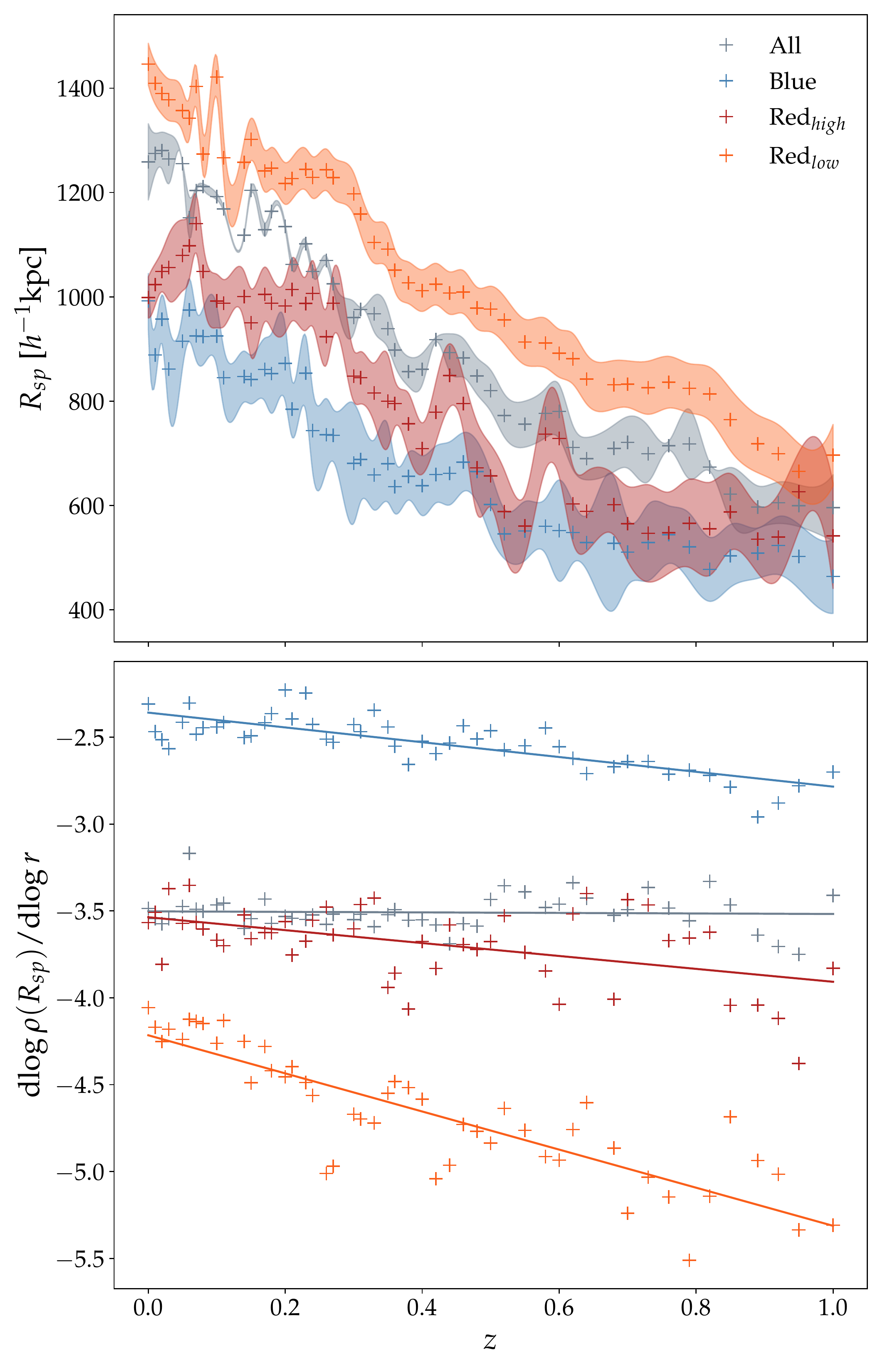}
    \caption{Evolution of the splashback radius over redshift. The upper panel displays the evolving value for $R_{sp}$ in comoving units (kpc/h) for the three color-mass populations and all galaxies in our sample between $z=0$ and $z=1$. The uncertainty in $R_{sp}$ is  represented as the shaded region around the lines for each population. The lower plot displays the  evolution of the minimum of the logarithmic slope of the galaxy number density profile. Linear fits  are included as solid lines. Note that at all times the $\redl$ population has the largest splashback radius. %, and increases consistently as time goes on. %Blue and $\redh$ have significant overlap at early times, but at around z=0.4, $\redh$ galaxies start to exhibit a larger splashback radius. The splashback feature becomes less steep over time or stays constant for all the color populations.
    }
    \label{fig:totalsplash}
\end{figure}

We note that the relative trend in the location of the minimum slope between different galaxy colors holds as a function of redshift (\autoref{fig:totalsplash}). 
Overall, the comoving radius of the minimum slope falls as a function of redshift for all populations. However, in units of $r/R_{200m}$, it remains fairly constant. In the bottom panel of \autoref{fig:totalsplash} we plot the depth of the slope-minimum as a function of redshift. The splashback feature remains at a constant depth for the All-galaxy profile but gets shallower at late times for all of the color-mass populations. This depth evolution is most extreme for the $\redl$ population. The shallow nature of $\redl$ splashback at late times is due to the higher density of $\redl$ galaxies outside the splashback radius that are in smaller groups before infall into the massive clusters considered here. The depth of splashback radius is thus sensitive to the density of galaxies in the neighbourhood of the cluster.

\textit{Summary:}  Red galaxies in a cluster are not a uniform population, but originate from three different channels. The massive galaxies that are centrals of accreted group mass halos have quenched mostly before infall and their splashback is shifted to a smaller radius compared to the total population. The low-mass Red galaxies, are i) either brought in by these groups, pre-processed previous to infall as can be seen in the large velocity clouds in the infall region of \autoref{fig:phase_space_colors}, or ii) are the remnants of Blue galaxies that have quenched in the cluster themselves. The Blue and $\redl$ galaxy outer slopes are therefore  probes of quenching within clusters.

\subsection{Comparison with Data}
\label{Sec:comparison_with_data}

\begin{figure*}[t!]
    \centering
    \includegraphics[width=8.5cm]{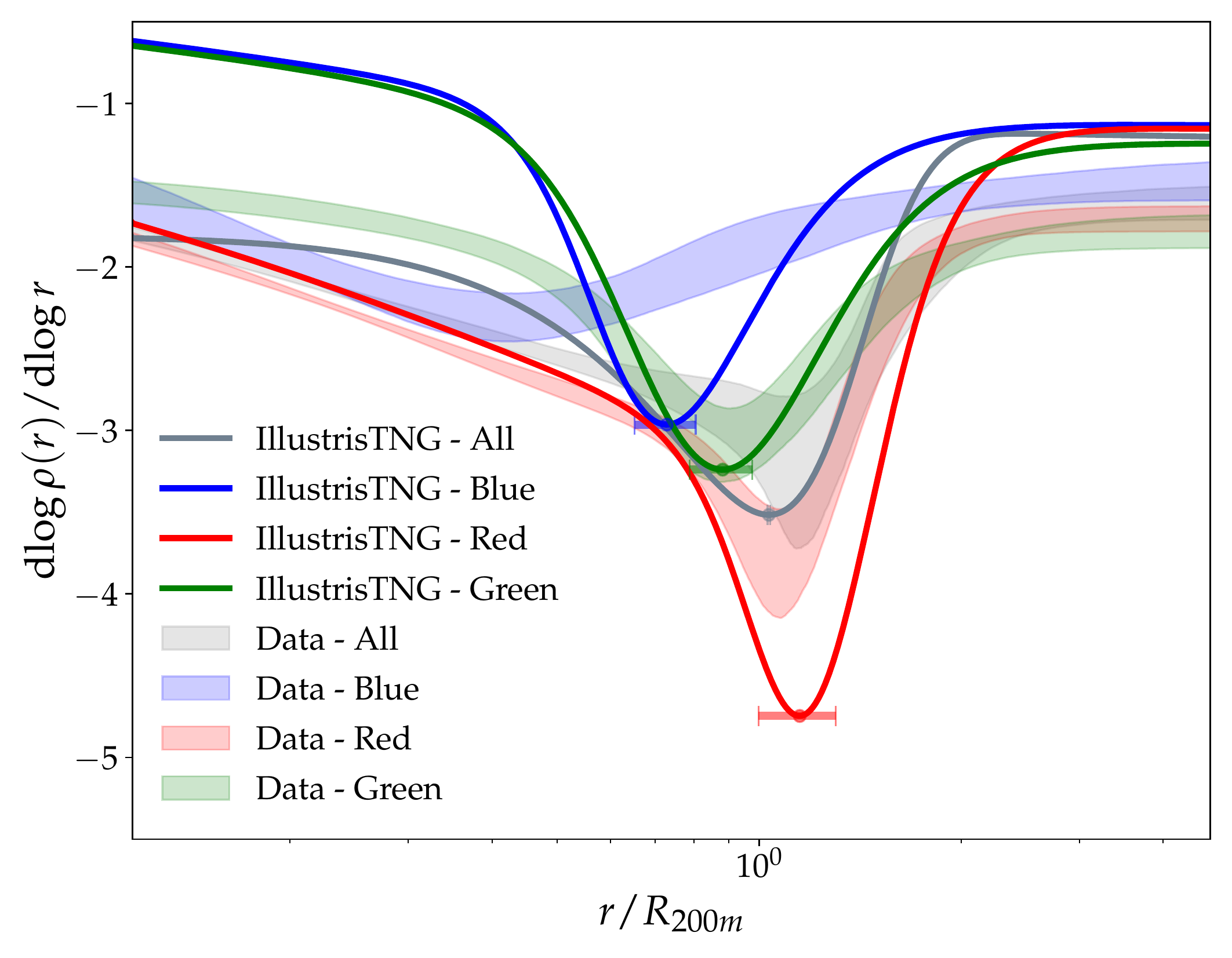}
    \includegraphics[width=8.5cm]{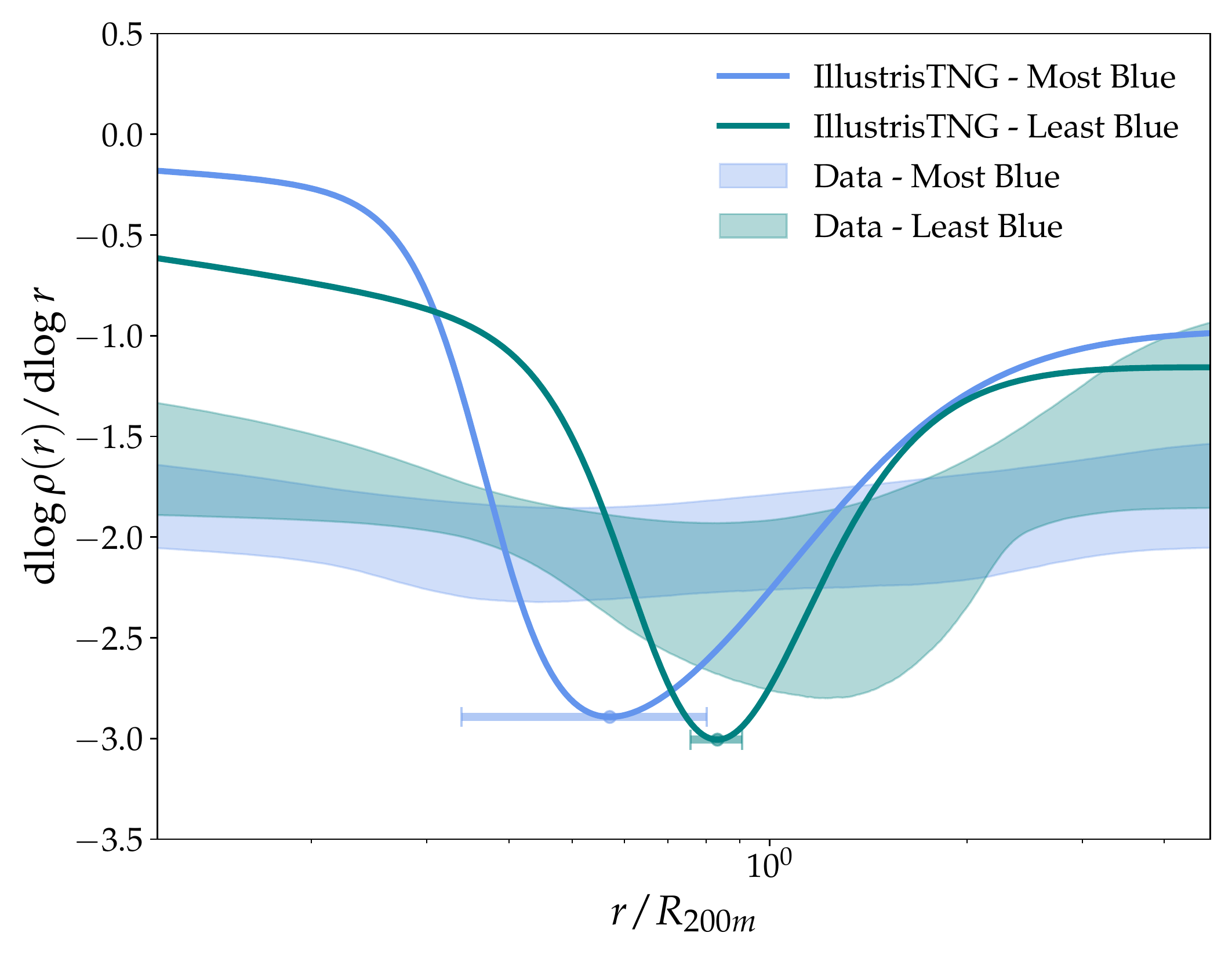}
        \caption{\textit{(Left)} Fitted slope of the  3d galaxy number density profiles split by color, comparing observational  data from \cite{adhikari20} to similar color and magnitude cuts  in IllustrisTNG. The shaded curves are measurements from data, and the solid curves are from IllustrisTNG. The location of the minimum slope of red and green galaxies in IllustrisTNG agree well with the data, while the blue galaxy population minimum slope looks different. %The minimum slope value of the red galaxy population is also significantly lower in IllustrisTNG compared to in data. \sa{I edited this out, because the depth relates to something subtle(things outside the cluster, and the data is averaged over 0.2-0.7)}
        We discuss further comparisons in the text \ref{Sec:comparison_with_data}. \textit{(Right)} Fitted slope for 3d galaxy number density profiles of Blue galaxies split into bluer and less blue sub-populations, comparing observational data to similar cuts made in IllustrisTNG.  IllustrisTNG exhibits steeper and more distinct slope minima at smaller radii than data, particularly for the most blue population. The inner regions are also significantly shallower in IllustrisTNG. (Note that the cluster mass threshold in IllustrisTNG is slightly lower than the threshold in data. This is in order to obtain good statistics in the simulation volume).}%we retain a host halo mass cut of $5\times 10^{13} h^{-1} M_{\odot}$ in order to have enough clusters and galaxies to fit, resulting in a lower mass cluster sample compared to the SPT data.)}

        %however the distribution of halo masses is not matched, and is lower than in ACT. We also select galaxies from all subhalos with an $i-$band magnitude greater than $19.7$. The data comparison color splits for IllustrisTNG Galaxies are defined as follows: Blue galaxies have $g-r<0.4$, Green galaxies have $0.4<g-r<0.55$, and Red galaxies have $g-r>0.55$. The shape of the profiles and location of the splashback feature is consistent between ACT and IllustrisTNG for both the Red and Green populations, however Red galaxies in IllustrisTNG have a deeper feature. However, the Blue populations are not consistent, with a very different shape for the inner profile. Illustris has a much deeper signal for splashback for Blue galaxies, whereas ACT data has little evidence for a splashback feature in Blue galaxies. \sa{discuss in text? we can also think more about what part to include here} \mb{I shortened the discussion}} 
        
    \label{fig:dataprofile}
\end{figure*}

In this section, we compare the galaxy number density profiles from IllustrisTNG with results from \cite{Shin19} and \cite{adhikari20}. Both these studies use the SZ-selected clusters and DES galaxies to measure the cluster-galaxy correlation function. The mean mass of the sample in \cite{adhikari20} is $M_{200 \rm m}=3\times 10^{14} h^{-1} M_{\odot}$. Because there are very few clusters in the small volume of IllustrisTNG we select a halo sample of $M_{200m} > 10^{14} h^{-1} M_\odot$ to compare with data. We select only subhalos with an $i-$band magnitude brighter than $-19.87$ and non-zero total mass. We then attempt to select three similar color populations by using only $g-r$ color, instead of color-space as in \cite{adhikari20}. For this purpose, we define blue galaxies as those with $g-r<0.4$, green as those with $0.4< g-r < 0.55$, and red as those with $g-r>0.55$. In the left panel of \autoref{fig:dataprofile}, we overplot the ACT results for galaxy splashback by color from \cite{adhikari20} with results from similar color cuts imposed on a similar sample from IllustrisTNG. 

We find that the splashback radius of the red IllustrisTNG profile is similar to the red data splashback radius, however the splashback signal in IllustrisTNG is significantly stronger, indicating a steeper fall off of the profile at the splashback radius. The shape of the log derivative is remarkably close, with the slopes in inner regions agreeing exceptionally well. Similarly, we note that the slope minimum for green galaxies agrees well with the data. However, we note that the green galaxies have a steeper profile than that observed in data within the halo. 

For the blue profiles, we see significant differences between IllustrisTNG and observed clusters. \citet{adhikari20} and \citet{Shin19} found no evidence of a feature at the splashback radius for blue galaxies, implying that blue galaxies do not survive to reach their first orbital apocenter --  the measured density profile remains at least as steep as the infall profile well within the splashback radius. There is a shallow minimum, thought to be associated with their quenching and ceasing to be classified as blue galaxies. 
In IllustrisTNG, we see a deeper slope-minimum and a much shallower inner profile compared to observed clusters. A combination of rapid quenching along mostly radial orbits, combined with a significant population of blue galaxies with more tangential orbits, could provide a possible explanation for these differences. A rapid quenching process of blue galaxies on radial orbits (in particular, if a large fraction of galaxies within a radius at some fraction of $R_{sp}$ undergo severe environmental quenching) would suppress their density, causing the slope of the number density profile in the inner regions to approach zero. Blue galaxies on non-radial orbits (such orbits have smaller apocenters than the most radial ones), would quench more slowly -- presumably after reaching apocenter -- and thus steepen the profile at radii slightly smaller than $R_{sp}$. 
Much of this discussion is supported by the phase space diagrams of \autoref{fig:phase_space_gas} which show the fractional gas mass loss of the three galaxy types. \textit{This suggests that the quenching of galaxies on radial orbits is more rapid in IllustrisTNG than in data, while those on more tangential orbits take a longer time to quench.} We also note that the effect of projection in going from 2d to 3d may play a role in comparing with observations. 

In order to further probe the differences between blue galaxies in IllustrisTNG and data, we follow the analysis of \cite{Shin19} by determining the splashback of two populations that result from the splitting of the blue population into evenly split lower and higher $g-r$ groups \citep{Shin19}. We do this as a further check of how color quenching is related to splashback and the time a galaxy has been inside the cluster. The right panel of \autoref{fig:dataprofile} shows the results of a comparison to SPT data. The SPT data uses a host halo sample with  $\langle M_{500c}\rangle = 3\times10^{14} h^{-1} M_{\odot}$. In this comparison, we revert to our fiducial host halo mass cut of $5\times 10^{13} h^{-1} M_{\odot}$ in order to obtain good statistics in the simulation volume. 
% \sa{Why do we use a different set of clusters for ACT and SPT?}. 
As in \cite{Shin19}, we find that the "most blue" galaxies display a shallower slope minimum, at a smaller radius than the "less blue" galaxies. However, the difference between the depth and location of the slope minima in IllustrisTNG is much less significant. Both blue populations in IllustrisTNG have more distinct slope minima and significantly shallower inner regions than in SPT data. 
% This aligns with the notion that the bluest galaxies are still not at first apocenter and have spent the least time in the cluster. However, the much shallower slope in Illustris implies that blues on most radial orbits do not reach pericenter, which is not true in data. 
Since no subset of the blue galaxies mirrors the slope of the profiles in data, we emphasize the argument that there are inherent differences between the blue galaxy population in IllustrisTNG and in data. 

\begin{figure*}
\centering
    \includegraphics[width = 5.95cm]{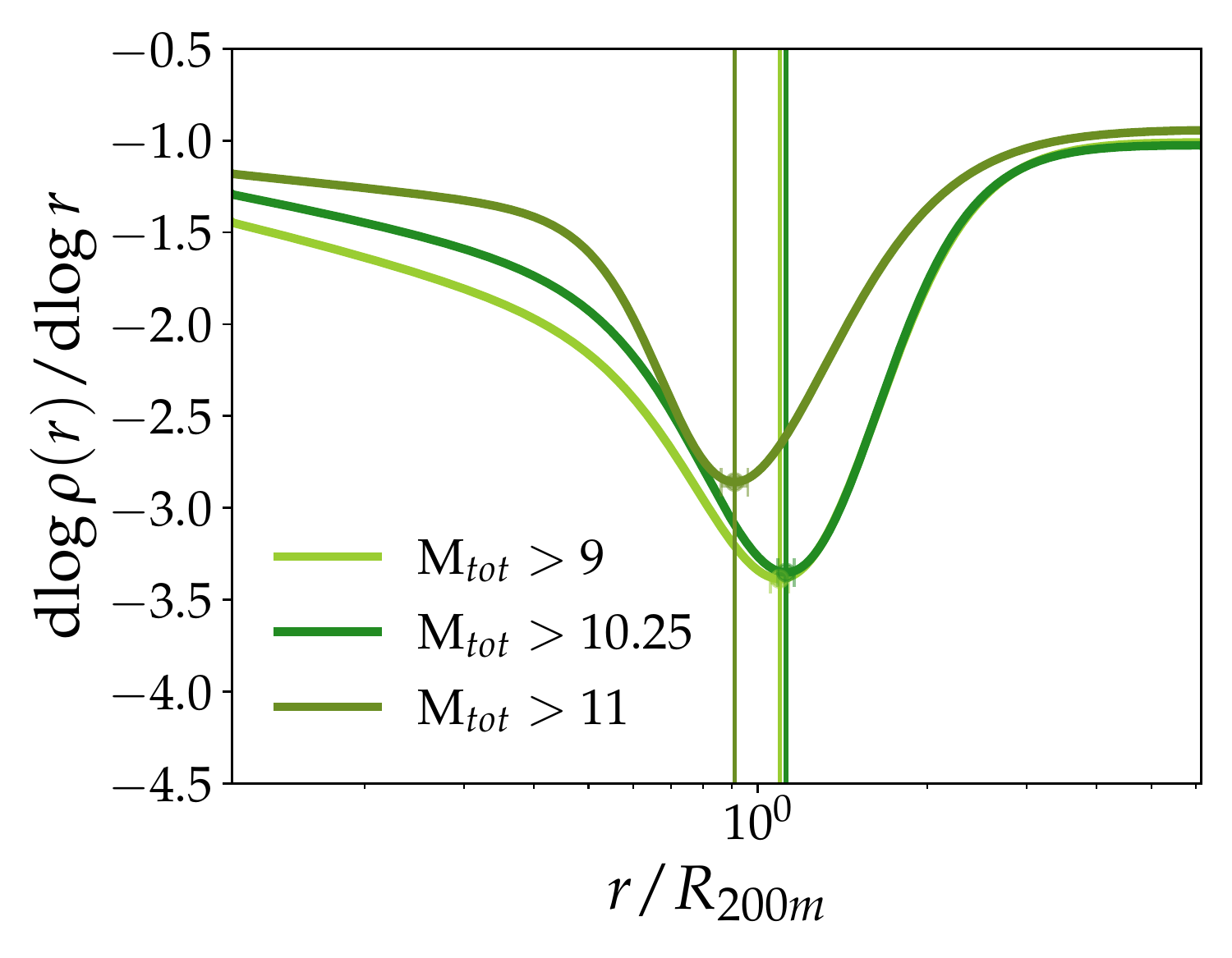}
    \includegraphics[width = 5.95cm]{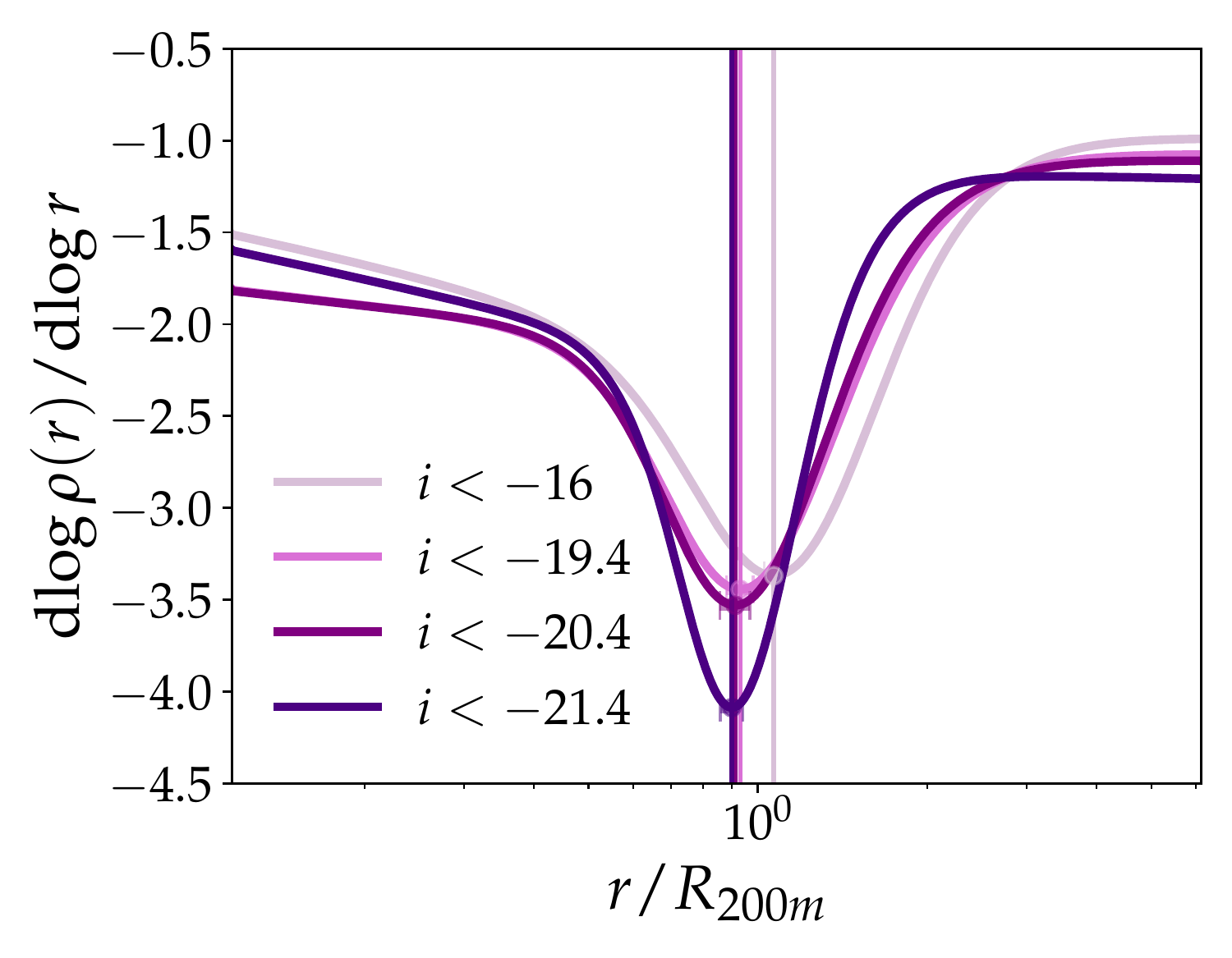}
    \includegraphics[width = 5.985cm]{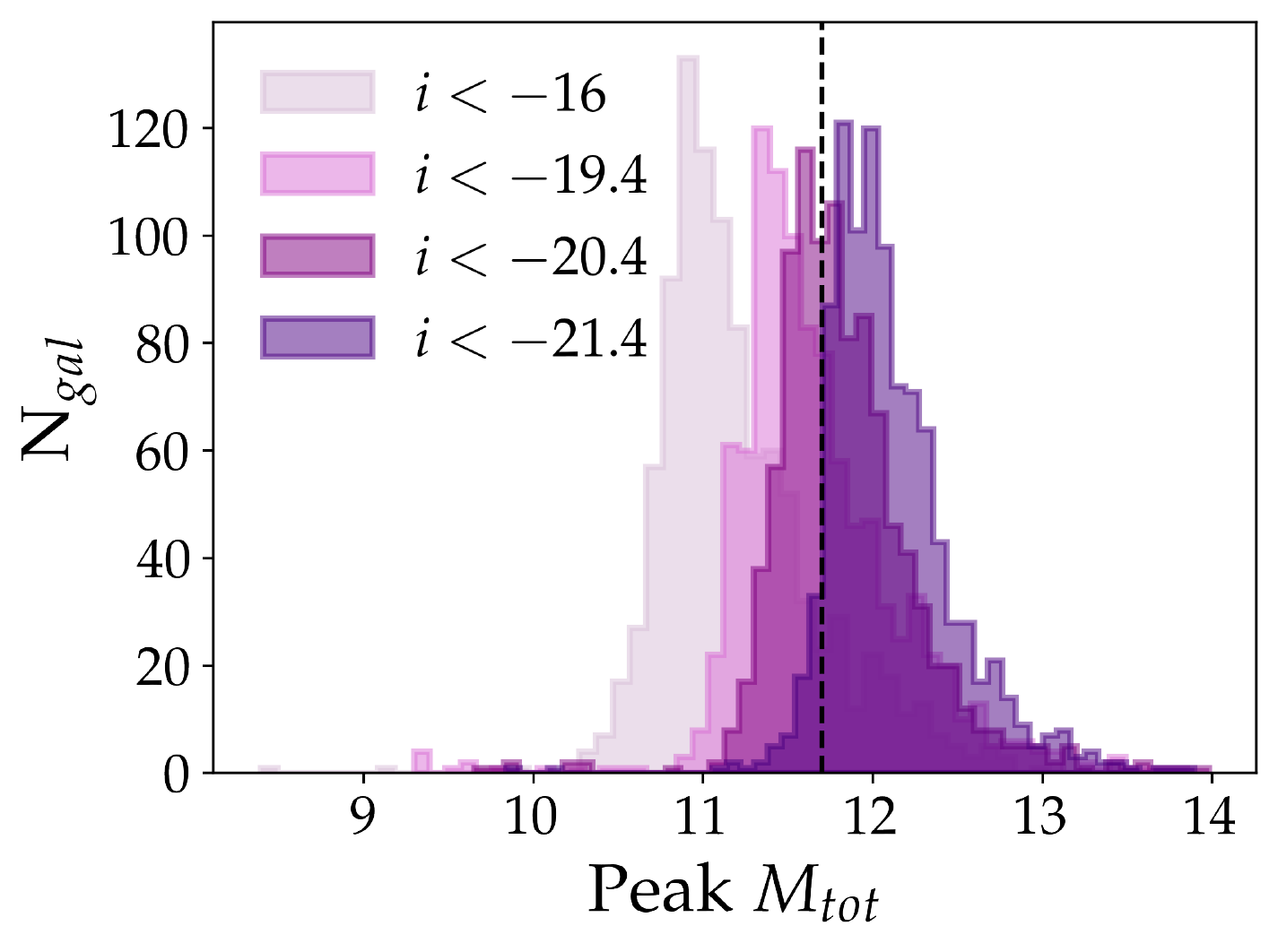}
    \caption{\textit{(Left)} Fitted slope of 3d number density profiles as a function of total mass (all galaxies split into three mass bins: greater than $10^9, 10^{10.25}, 10^{11} h^{-1} M_\odot$). 
    % The most encompassing mass bin, $M>10^9 h^{-1} M_\odot$, has an $R_{sp}$ between the two larger bins, with $M>10^{10.25} h^{-1} M_\odot$ having an insignificantly higher $R_{sp}$, and the highest mass bin at $M>10^{11} h^{-1} M_\odot$ having a notably lower splashback. 
    The lower splashback radius of the highest mass bin is consistent with the effects of dynamical friction. \textit{(Middle)} Fitted slopes of all galaxies in our sample split into four bins based on $i-$band magnitude ($i<-16$, $i<-19.4$, $i<-20.4$, $i<-21.4$). The $i<-16$ profile can be considered similar to the All-galaxy profile in our usual sample. The other three magnitude bins have very similar values for splashback. Thus splitting galaxies by their absolute magnitude does not seem show effects of dynamical friction. \textit{(Right)} Peak $M_{tot}$ distributions for the same $i$-band splits used in the middle plot. The black dashed line indicates $M_{tot}$ = $0.01 M_{host,min}$ where $M_{host,min}=5\times10^{13} h^{-1} M_{\odot}$ is the minimum total host mass. See text for further discussion.  }
    \label{fig:dynfric}
\end{figure*}

We note also that the slope profile for the full galaxy sample (grey solid line in the left panel of \autoref{fig:dataprofile}) agrees statistically with data (grey shaded) near the splashback region. This was also shown in \cite{Shin_2021}. Both in the outer and inner regions there are however significant differences. In particular in the outer region the slope is significantly shallower in IllustrisTNG compared to data. As the slope in the outer region can be thought to be set essentially by free fall onto a massive cluster \citep{Baxter:2017csy}, this difference could possibly arise due to differences in the overall shapes of the two samples \citep{2011ApJ...734..100L}. More massive clusters (like those in data) are more elliptical than their lower mass counterparts and ellipticity can significantly affect the outer slopes in the direction seen here \citep{2011ApJ...734..100L}. Projection effects in data may also play a role. We defer a thorough investigation of this effect to future work. The differences in the inner region may be related to differences in galaxy disruption between the simulation and data.

\label{Sec:DynFricSec}
\subsection{Splashback and Galaxy Luminosity: Effects of Dynamical Friction and Self-friction}

In this section, we briefly explore the effects of changing the absolute magnitude selection of the galaxy sample on the  splashback feature. As discussed above, one of the potential explanations for the shift to lower splashback radii in high mass $\redh$ galaxies in  our fiducial host cluster sample is dynamical friction. Dynamical friction, through an effective drag force, can slow massive galaxies down and cause second turnaround to occur at a smaller radius. In particular, dynamical friction is known to bias the splashback radius w.r.t particles when the galaxies live in subhalos with total mass greater than 0.01 times the host mass \citep{More:2016vgs, Adhikari:2016gjw}.

Firstly, we split our galaxy sample into three cumulative total mass bins and fit number density profiles in the left panel of \autoref{fig:dynfric}. For this analysis, we revert back to our fiducial sample of clusters with $M_{\rm 200m}>5\times10^{13} h^{-1} M_{\odot}$. The two lower mass bins do not demonstrate a shift in slope minimum but the highest mass bin with $M_{tot} > 10^{11}$ displays a visible shift towards a lower splashback radius. These galaxies live in subhalos that are around 0.01 the mean host mass, and a shift is therefore expected. Note that the total mass in IllustrisTNG is the current subhalo mass, which has presumably lost a fraction of its mass since infall. Therefore, its pre-infall mass is expected to be even higher.

To compare with observational work in \cite{Chang:2017hjt}, we conduct a similar analysis but split the galaxy sample using absolute magnitude in the $i-$band in the middle panel of \autoref{fig:dynfric}.  \cite{Chang:2017hjt} did not find significant evidence for movement of the splashback radius with galaxy magnitude, even though abundance matching of their galaxy sample to CDM-only simulations implied that such a movement should be seen. 

Curiously, even though the RedMaPPer \citep{Rykoff2014} optically selected cluster sample in \cite{Chang:2017hjt} is expected to suffer from systematics, we see a similar trend in IllustrisTNG. Our magnitude binned samples show virtually no difference in splashback radii with $i < -19.4$ and $i < -20.4$ bins displaying almost identical splashback features while our highest magnitude bin, $i < -21.4$ has a significantly deeper splashback. We include one sample with $i < -16$ which can be considered as almost all subhalos (and therefore dominated by the lower mass ones), which has a higher splashback radius than the other three bins and is expected to be comparable to the splashback in particles as seen in Figure 6 of \cite{Shin_2021}. We note that the splashback in total mass and that of the faintest galaxy sample is shifted to a larger distance with respect to the brighter galaxies. 

In the rightmost panel of \autoref{fig:dynfric}, we plot the peak mass distribution of the galaxies in the different magnitude bins. We select a sample of 1000 galaxies from our subhalo sample randomly from within each magnitude for a total of 4000 galaxies, then using the LHaloTree, we track their total mass through time and evaluate their peak masses before infall. The black dashed line is at $0.01 M_{\rm host,min}$ or $5\times10^{11} h^{-1} M_{\odot}$. As is observed, there is significant scatter in the peak total mass in each sample with a magnitude threshold. Therefore, it is possible that all galaxies with magnitudes greater than $M_i<-19$ show effects of dynamical friction. Only a small fraction of the $M_i<-16$ samples has masses above $0.01 M_{\rm host,min}$. While dynamical friction may explain the shift between the overall particle splashback and galaxies, it is still unclear why the galaxy splashback remains static between the different brighter galaxy samples.

It is important to emphasize another possible cause for the shift to smaller splashback radii: the effect of dynamical self-friction. Dynamical self-friction is an effect that can cause a significant reduction in the orbital angular momenta of subhalos due to the torque exerted by material stripped from themselves. It can potentially dominate over dynamical friction along more radial orbits and close to pericentric passage \citep{Miller_2020}. Work done by \cite{Miller_2020} found that in simulations, massive subhalos in particular experience this self-friction. It causes subhalos to spiral towards the centers of host clusters, an effect which, like dynamical friction, could potentially reduce the apparent splashback radius of an affected galaxy population. A quantitative study of the interactions of dynamical friction and self-friction as a function of halo mass, splashback, and infalling groups in IllustrisTNG would be able to shed more light on the relative impact of this effect.

Note that \cite{neil21} also found that the splashback radius was around 12\% lower in galaxy number density profiles as compared to dark matter profiles for a cluster sample between $10^{13}$ and $10^{13.5} M_{\odot}$. They attributed this difference to dynamical friction. We see a difference of similar degree in our absolute magnitude split samples between the $i < -16$ and the higher magnitude bins, implying that friction effects must be taken into account when dealing with cluster samples that are closer to group sized halos.

\section{Conclusion}
\label{Sec:Conclusion}

We have studied the evolution of galaxies in galaxy clusters in the IllustrisTNG 300-1 simulation and connected them to observable signatures in their number density profiles, phase space distributions, and splashback radius. Our study has two additional goals: 1) to understand IllustrisTNG's galaxy populations using these observables and to compare them with observations from DES, and 2) to use the phase space, color, and luminosity distributions of galaxies to test our methodology of connecting quenching during galaxy infall  into clusters with their radial profiles (see \cite{adhikari20}).  Hydrodynamical simulations offer a useful test of the methodology since quenching is implemented in a  physical way in contrast to N-body simulations. Finally, we also explore the effect of dynamical processes like dynamical friction and self friction on the distribution of galaxies. 

We use Gaussian mixture models in the space of color and total mass to identify three distinct galaxy populations: Blue, $\redl$ and $\redh$. We compare these with the populations studied in previous work. We explore how the different evolutionary channels for the three types of galaxies leave signatures on the overall color distribution of the galaxies in the universe and in the environment of galaxy clusters. We also make preliminary comparisons with existing data on galaxies observed with the DES optical survey around massive clusters identified via their SZ signal in the ACT and SPT surveys.
%Observing our galaxy populations evolving in color-mass space over time reveals the distinct pathways taken by galaxies that reside in the different populations. 
We find that, on average, $\redh$ galaxies originate as Blue galaxies that increase significantly in mass and then undergo a fast quenching process at a certain mass threshold, consistent with ``mass quenching'' which typically occurs due to AGN feedback while these galaxies infall as centrals of groups. $\redl$ galaxies originate as Blue galaxies that redden with time in the dense environments of either infalling groups or the clusters. These two different paths reflect the mass and environmental quenching experienced by high and low mass Red galaxies, respectively. We study in detail how these galaxies occupy the phase space of cluster halos.

%While clusters are known to have an overdensity of red galaxies, these galaxies undergo three different evolutionary processes which have also been discussed in \citet{Donnari_2020}. 
We also study their impact on the density profile of clusters, which is observed in projection in multi-color imaging surveys. Massive Red galaxies, ($\redh{}$), falling in as centers of groups with large subhalo masses, tend to sink to the center of the cluster due to the combined effects of dynamical friction and self-friction. These galaxies therefore appear to have a significantly smaller splashback radius compared to the full galaxy population. In addition, $\redh$ galaxies, which are predominantly already red \textit{before} infall, lose all of their gas mass within the virialized region of the cluster. 
%%A cross-correlation of clusters with massive red galaxies, should help constrain what fraction of such galaxies exist in observed clusters. 
%The low-mass $\redl$ galaxies on the other hand either fall in as formerly quenched satellites of groups, or originate within the cluster as evolved Blue galaxies that redden as their star-formation rate decays with time. The fraction of galaxies that fall in as quenched satellites increases at late times (see \autoref{fig:phase_space_3z_weighted}). 

The Blue galaxies in the vicinity of a cluster primarily live in the infall stream. They show a clear loss of gas fraction as they traverse the phase space within the cluster halo (see Fig. \ref{fig:phase_space_gas}). In IllustrisTNG, there are very few Blue galaxies with radial orbits that cross pericenter and remain star-forming -- we find that the highest positive radial velocity envelope in phase space is missing  Blue galaxies which have converted into low-mass Red galaxies. %These galaxies therefore do not reach the farthest distances in their apocenter. 
Their phase space boundary, between multistreaming and infall, is therefore also at a smaller radius than the splashback boundary of the halo. Therefore, the farthest boundary of the halo is made of $\redl$ galaxies. Thus, low-mass $\redl$ galaxies sample the complete phase space of dark matter in a halo (barring disruption of a small fraction in the innermost region). In other words, all possible phase space locations of dark matter should be most closely sampled by the low-mass $\redl$ galaxies. A significant fraction of the $\redl{}$ galaxies are also brought in as part of groups, this fraction increases at late times.

While the above discussion focused on the different color populations, we emphasize that the total galaxy profile closely traces that of the dark matter (when the galaxy magnitude threshold is not significantly affected by dynamical friction). The quenching induced changes in galaxy color do not affect the shape of the total profile, keeping the total number of galaxies conserved. We note that the agreement of the total galaxy profile with the matter profile in \cite{Shin_2021} is consistent with the idea that processes such as tidal disruption do not destroy a significant fraction of the galaxies in the sample. This finding has two direct implications: that if galaxies follow dark matter closely, they can be used to study halo properties, and deviations from the dark matter distributions of galaxy sub-populations help us study evolutionary processes.  %studying different subselections of galaxy populations will tell us about their evolutionimportant to understand disruption models both in hydrodynamic simulations and in constraining models of artificial disruption in N-body simulations. $This finding is consistent with the observational studies of Shin et al (2021) and others.

Comparing DES galaxies around ACT and SPT clusters to the simulations revealed subtle differences between the galaxy populations (see \autoref{fig:dataprofile}). In particular, while the red and green galaxy splashback and radial profiles agree fairly well, the blue galaxies behave significantly differently from simulations. We find that the observational data prefers a scenario where the blue galaxy profile remains close to a power law from the infall regime down to the inner halo, with a shallow minimum inside the halo, implying that some blue galaxies cross pericenter and quench before apocenter. However in IllustrisTNG the profile flattens: the shallow slope in the inner regions shows that blue galaxies are either getting disrupted or completely quenched by pericenter. In addition, the sharper feature in the outskirts in IllustrisTNG is formed by a significant fraction of galaxies that are on higher angular momentum orbits. Therefore, it appears that for the simulation clusters, the quenching time scale is shorter than that in observational data. 

Our work suggests that the evolutionary histories of galaxies through their orbits around a halo leave distinct signatures on their phase space distribution. Observationally these signatures are encoded in the shape of radial number density profiles and also the projected velocity distributions. In  particular, we note that the distribution of blue galaxies may be most informative about the quenching timescales which likely depend both on the orbits and infall times. In future work we hope to constrain the orbit and infall time distribution of galaxies of different population using the rich data that is available. We note that in spectroscopic surveys we can jointly model the velocity and spatial distributions to constrain models of quenching and disruption. The local slope of the number density distribution alone also encodes much of the phase space information, making it a powerful probe of galaxy evolution in photometric surveys.  Current surveys can measure these distribution out to $z=1$ with high  precision. Comparison to the dark matter distribution  \citep{Shin_2021}, or the gas distribution \citep{2016arXiv161002644Z, Baxter:2021tjr, Anbajagane:2021bnx} can help us distinguish between different intra-cluster processes that are relevant for galaxy evolution, help constrain models in hydrodynamic simulations, and  fine tune our understanding of the galaxy-halo connection in clusters.
\vspace{0.5cm}
\section*{Acknowledgements}

We thank Dhayaa Anbajagane for comments on an early draft of the paper, and Chihway Chang, Eric Baxter and Shivam Pandey for insightful discussions. We also thank Stephanie O'Neil for useful discussion and comments. SA was supported by a grant from the U.S. Department of Energy (DOE), and the DOE Office of Science Distinguished Scientist Fellow Program. We thank the IllustrisTNG collaboration for providing free access to the data used in this work, which can be accessed at \url{www.tng-project.org}.

%The inner slope of the radial number density profile of blue galaxies remains as steep as the infall term, however in simulations the inner slopes are much shallower. The slope minimum is also much deeper and pronounced in the Illustris simulation. These differences arise from the inherently faster quenching timescales for galaxies in the Illustris simulation that live on radial orbits. The data appears to prefer a slower quenching timescale, allowing blue galaxies to cross pericenter, even on the most radial orbits. These differences can help constrain models of feedback and star-formation evolution in hydrodynamic simulations.
\bibliography{references}

\appendix
\section{Hydrogen Abundance}
\label{app:Habundance}

\begin{figure*}[h]
    \centering
    \includegraphics[width=18cm]{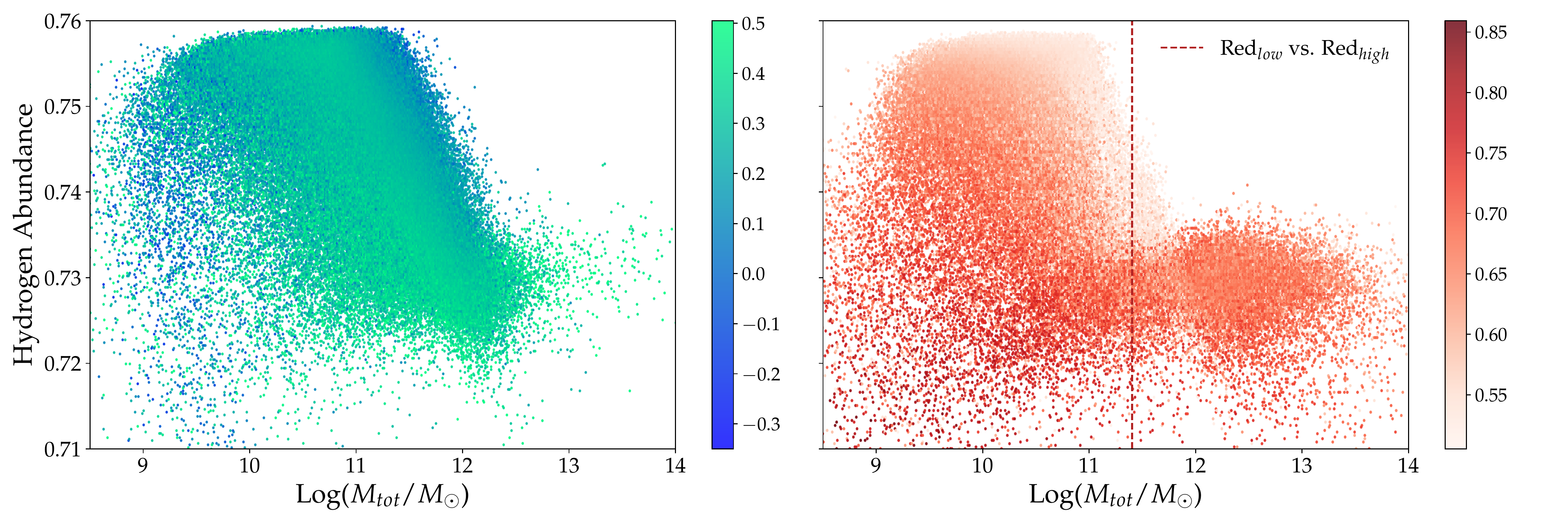}
    \caption{Hydrogen abundance of galaxies versus total mass, weighted by $g - r$ color. The left plot considers Blue galaxies only, while the left plot shows all Red galaxies. The dashed red line in the right plot indicates the $z = 0.52$ cut defined in total mass between $\redl$ and $\redh$ galaxies. The bluer Blue galaxies are bifurcated between a low mass, lower Hydrogen abundance region, and a higher mass, higher abundance region, in between which are redder Blue galaxies. In the right panel, the vertical line indicating the split between $\redh$, $\redl$ also splits two distinct populations in abundance vs. mass, such that $\redh$ galaxies have a lower Hydrogen abundance than the $\redl$ population. }
    \label{fig:H_abund}
\end{figure*}

We briefly examine the distributions of the three color-mass populations  in hydrogen abundance to further probe their diverging properties. \autoref{fig:H_abund} shows the hydrogen abundance of galaxies at $z = 0.52$ versus mass, weighted by the $g - r$ color of the galaxy. Blue galaxies display a complicated relationship between hydrogen abundance, mass, and color. Generally, the higher the mass, the lower the hydrogen abundance. The \textit{bluer} (lower $g - r$) Blue galaxies are located at the extremes of the hydrogen to mass relationship. The most blue galaxies either have much higher proportions of hydrogen relative to mass compared to the average blue galaxy, or a much lower ratio. For example, along a line of constant mass at $10^{11} h^{-1}M_\odot$, the highest and lowest ends are occupied by the most blue galaxies. Meanwhile, redder Blue galaxies tend towards the median, rather than occupying one or the other extreme. This relation further suggests that galaxies which are initially blue quench into two different populations based on their initial conditions, which can be visibly distinguished in \autoref{fig:H_abund}. However, further work is necessary to determine how Blue galaxies in IllustrisTNG actually evolve in this hydrogen abundance - mass space.

The split in total mass between $\redl$ and $\redh$ is noted in the plot as the dashed red line. This split clearly separates Red galaxies with distinct behavior in hydrogen abundance - mass space. $\redl$ galaxies display a similar relationship between hydrogen abundance and mass as Blue galaxies, the higher the mass, the lower the hydrogen abundance. This correspondence between the properties of Blue and $\redl$ galaxies reflects their behavior on the SFMS. $\redh$ galaxies, on the other hand, are clustered at a small range of low hydrogen abundances with little to no relationship between mass and abundance. This behavior could align with the idea of $\redh$ as the products of an internal mass-quenching process linked to AGN feedback as that feedback process has been linked to a reduction of the neutral hydrogen content of halos in hydrodynamical simulations \citep{Villaescusa_Navarro_2016}.   

\begin{figure*} [h]
    \includegraphics[width=18cm]{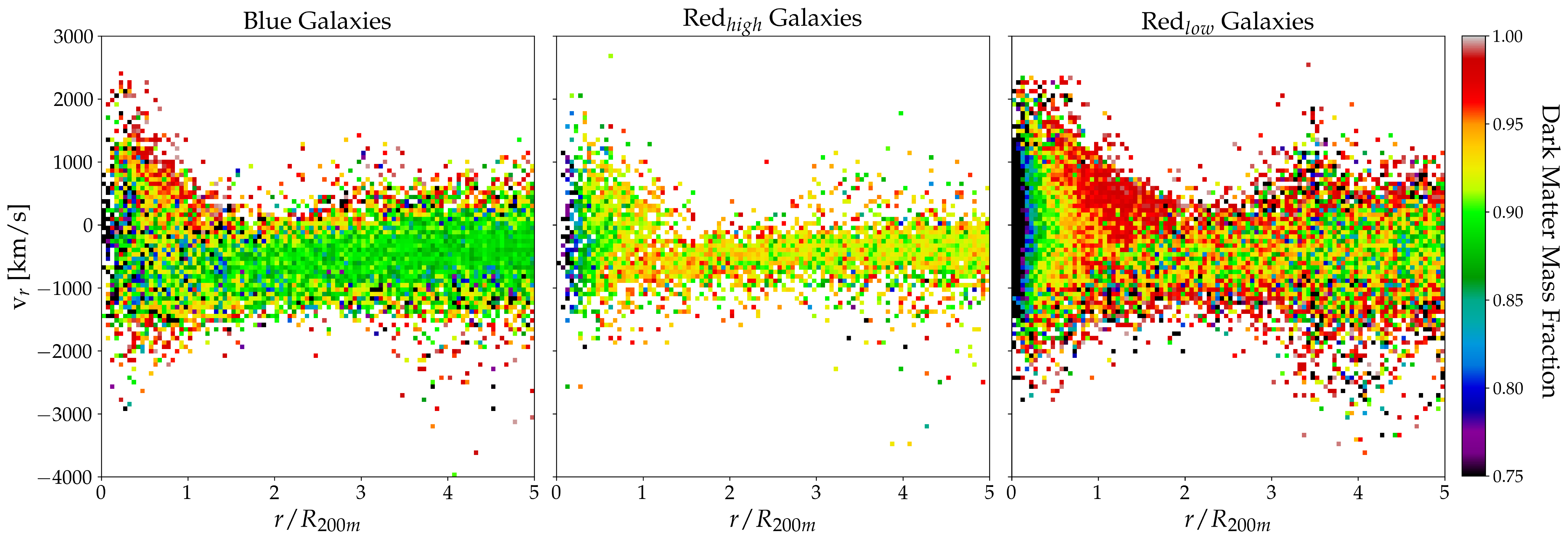}
    \centering 
    \caption{Stacked phase space for three color-mass populations, where each panel is colored by the mean ratio of mass of dark matter to total mass of the galaxies in a given bin. Note that the density of galaxies in each histogram bin is not represented in this plot. Blue galaxies have relatively constant ratios of dark matter, with higher ratios at higher velocities. $\redh$ galaxies have decreasing ratios of dark matter to total matter as they enter the multistreaming region and fall into the cluster. $\redl$ galaxies  have varied amounts of dark matter in the infall stream, however have a higher dark matter ratio between pericenter and splashback than in infall, before rapidly decreasing in cluster interiors.}
    \label{fig:phase_space_DM}
\end{figure*}
\begin{figure*} [h]
    \includegraphics[width=18cm]{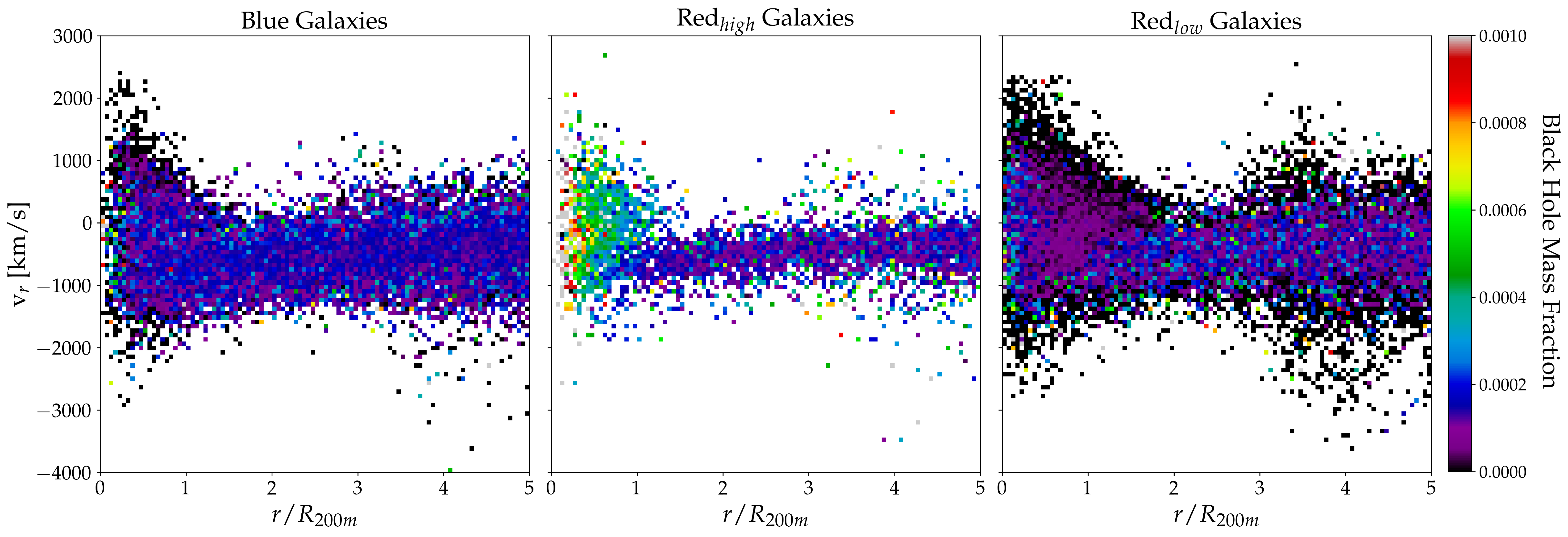}
    \centering 
    \caption{Stacked phase space for three color-mass populations, where each panel is colored by the mean ratio of black hole mass to total mass of the galaxies in a given bin. Note that the density of galaxies in each histogram bin is not represented in this plot. Blue galaxies and $\redl$ galaxies have similar black hole ratios, with higher velocity galaxies have little to no mass contribution from black holes. $\redh$ galaxies have higher black hole mass ratios towards cluster interiors.}
    \label{fig:phase_space_BH}
\end{figure*}
\section{Distribution of Dark Matter and Black Hole mass in Phase Space}
\label{app:phasespace}
We extend the discussion from \autoref{Sec:matterdist} by examining the black hole and dark matter ratio distributions in phase space for the three color-mass populations defined in \autoref{Sec:Color-mass}. In \autoref{fig:phase_space_DM}, we show the stacked phase space with each bin weighted by the mean dark matter mass ratio (DM/total) for Blue, $\redh$, and $\redl$ galaxies. All three populations exhibit a trend of decreasing dark matter ratio towards the cluster center, demonstrating dark matter stripping. Previous studies in the Illustris simulation have shown that dark matter stripping begins at around $1.5 \cdot R_{vir}$ \citep{niemiec2019}, which is also where we see the infall stream of phase space transition into the multistreaming region. However, we overall find a slightly more complicated picture for dark matter stripping, as $\redl$ galaxies and Blue galaxies at high radial velocities and in the multistreaming region show higher dark matter mass fractions than galaxies in the infall stream. The initial peak in the $\redl$ curve in \autoref{fig:Mdm_fraction_curve} has a corresponding region visible as a red (high dark matter ratio) triangular area in the Red$_{low}$ DM fraction plot of \autoref{fig:phase_space_DM}, likely created by quenching of blue galaxies. The distinct higher dark matter ratio group of blue galaxies at higher radii reinforces the arguments made in \autoref{Sec:comparison_with_data} that there are significant differences in the quenching pathways of blue galaxies on radial and tangential orbits, which in turn impact the galaxy number density density profiles. In \autoref{fig:phase_space_BH} we similarly weigh each bin in phase by the black hole mass fraction for each of the three populations. We find further evidence for the distinct properties and quenching pathways taken by the two Red populations in their black hole mass fractions. $\redh$ and Blue galaxies both display higher black hole mass fractions than $\redl$ galaxies. This distinction between $\redh$ and $\redl$ galaxies supports the notion that $\redh$ galaxies are products of mass quenching by AGN feedback rather than environmental quenching. The higher radial velocity sections of the $\redl$ phase space have no black hole mass, further suggesting that they are undergoing infall as parts of smaller clusters. 

% \begin{figure*}
%     \includegraphics[width=18cm]{phase_space/Phase_space_subm_66.png}
%     \centering 
%     \caption{Stacked phase space for three color-mass populations, where each panel is colored by the mean total mass in a given bin. Note that the density of galaxies in each histogram bin is not represented in this plot. Blue galaxies have slightly higher masses in the infall stream at low velocities than at higher velocities or in the multistreaming region. $\redh$ galaxies seem to lose significant amounts of total mass as they fall towards cluster centers. $\redl$ galaxies however gave the highest total mass between pericenter and splashback, not in the infall region. \td{can rerun and maybe change colorbars etc but kept as a placeholder here now. Could also just keep central panel??} \sa{should we show only the third panel and combine with Fig. 8?}}
%     	  \label{fig:phase_space_subm}
% \end{figure*}

\section{Line of Sight velocity distributions}
\label{app:los}

We study the line of sight (LOS) velocity distribution of the three populations as a function of radius from the cluster center. The LOS velocity dispersion is also be a promising probe of galaxy evolution, particularly in spectroscopic surveys. They are also often used for mass callibration of clusters. We assume a flat sky approximation and use the $z$ component of the velocity of the galaxy w.r.t to the cluster center as the LOS direction such that,

\begin{equation}
    \Delta v_{\rm los}=\frac{v_{\rm gal}-v_{\rm cluster}}{1+z_c}
\end{equation}

\begin{figure*} [h]
    \includegraphics[width=18cm]{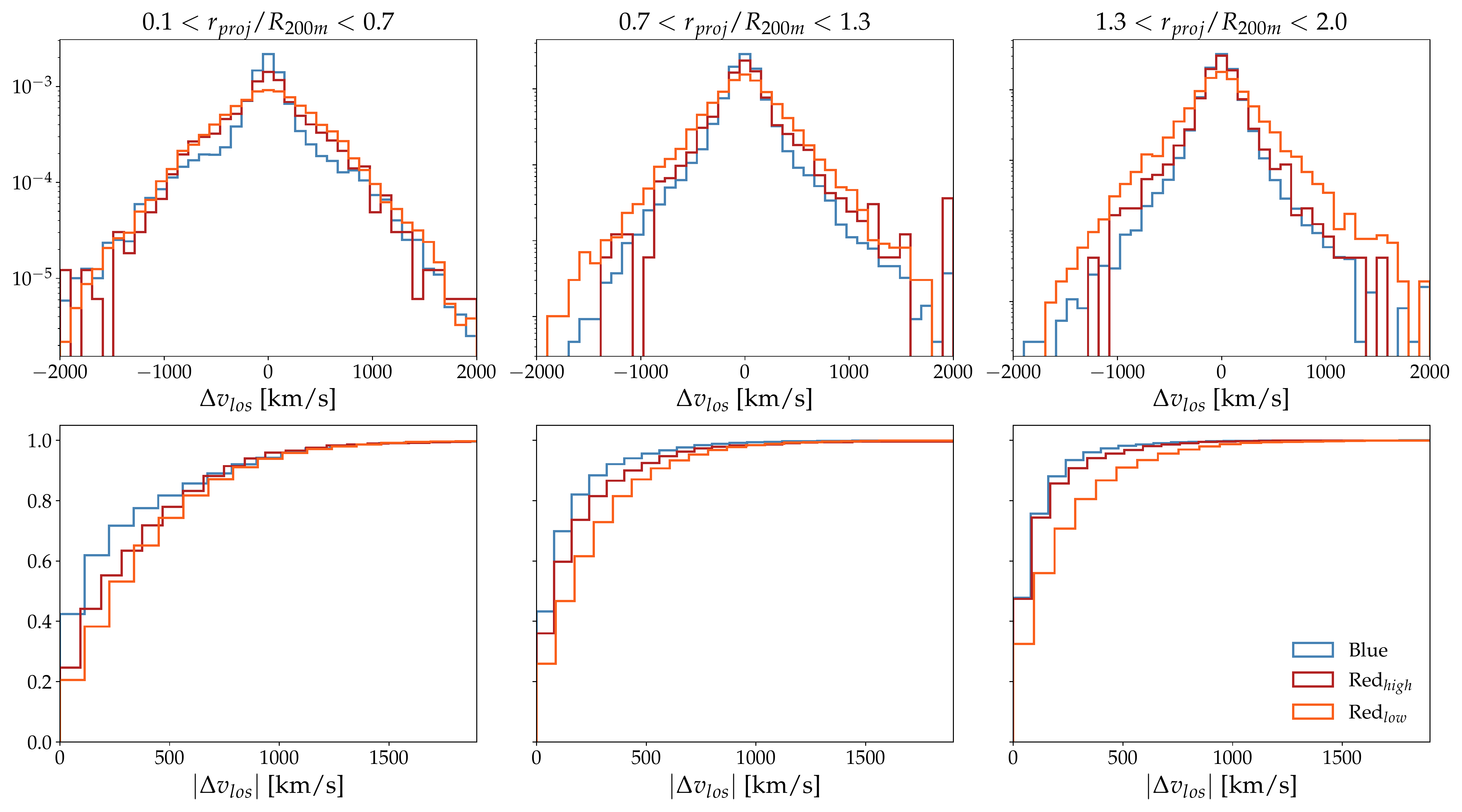}
    \centering 
    % \caption{Top series of panels: vlos distribution, bottom series of plots, vlos CDF.}
    \caption{The line of sight velocity (LOS) distribution as a function of galaxy color in three projected radial bins. The top panel shows the PDF and the bottom panel shows the CDF. The LOS distribution encodes signatures of galaxy evolution as discussed in the text.}
    \label{fig:los_dist}
\end{figure*}

We measure the LOS distributions for clusters at $z=0.52$, in three bins of projected radii shown in \ref{fig:los_dist}. The three bins were chosen to represent the innermost regions of the cluster, near the splashback radius and in the infall region. In the infall regime (rightmost panel), $\redl{}$ galaxies have strong tails, and the highest LOS velocity dispersion, this is because a significant fraction of them  they fall in as parts of groups. Blue galaxies are primarily tracing the host clusters potential.  In the innermost regions of the cluster, the blue galaxies and to some extent the $\redh{}$ galaxies show an LOS distribution that appears to be composed of two distinct components, with a small and large dispersion. The low dispersion component peaked at $\Delta v_{\rm los}=0$, is from galaxies on tangential orbits that have low LOS velocities at small projected radius. As discussed in the sections above, IllustrisTNG quenches radial blue galaxies, therefore a large fraction of galaxies that are blue will show this two-component LOS distribution. 

%\bibliography{references}
%\begin{thebibliography}{99}
%\bibitem{shin19} Shin et. al. Measurement of the Splashback Feature around SZ-selected Galaxy Clusters with DES, SPT and ACT
%\end{thebibliography} 

\end{document}